\documentclass[12pt,showpacs,amsmath,amssymb]{revtex4}
\usepackage{graphicx}
\usepackage{bm}

\def\br{ \bm{r} }
\def\bk{ \bm{k} }

\def\bp{ \bm{p} }

\def\bP{ \bm{P} }
\def\hbp{ \hat{\bm{p}} }
\def\bro{ \bm{\rho} }
\def\bgam{ \bm{\gamma} }
\def\bfg{ \bm{g} }
\def\bA{ \bm{A} }

\def\bE{ \bm{E} }
\def\bB{ \bm{B} }
\def\nablap{ \bm{\nabla}_{\bP} }

\def\sign{ \mathrm{sign} }
\def\tr{ \,\mathrm{tr}\, }

\begin{document}
\title{Spin-orbit coupling and semiclassical electron dynamics in noncentrosymmetric metals}

\author{K. V. Samokhin\footnote{E-mail: kirill.samokhin@brocku.ca}}
\affiliation{Department of Physics, Brock University, St. Catharines, Ontario L2S 3A1, Canada}


\begin{abstract}
Spin-orbit coupling of electrons with the crystal lattice plays a
crucial role in materials without inversion symmetry, lifting
spin degeneracy of the Bloch states and endowing the resulting
nondegenerate bands with complex spin textures and topologically
nontrivial wavefunctions. We present a detailed symmetry-based
analysis of the spin-orbit coupling and the band degeneracies in 
noncentrosymmetric metals. We systematically derive the semiclassical 
equations of motion for fermionic quasiparticles near the Fermi surface, 
taking into account both the spin-orbit coupling and the Zeeman
interaction with an applied magnetic field. Some of the
lowest-order quantum corrections to the equations of motions can
be expressed in terms of a fictitious ``magnetic field'' in the
momentum space, which is related to the Berry curvature of the
band wavefunctions. The band degeneracy points or lines serve as
sources of a topologically nontrivial Berry curvature. We discuss
the observable effects of the wavefunction topology, focusing, in
particular, on the modifications to the Lifshitz-Onsager
semiclassical quantization condition and the de Haas-van Alphen
effect in noncentrosymmetric metals.
\end{abstract}


\maketitle

\newpage
\section{Introduction}
\label{sec: Intro}

Majority of metals have a center of inversion in their crystal
lattice. Recently, however, there has been growing interest, both
experimental and theoretical, in the properties of
\textit{noncentrosymmetric} metals, driven mostly by their unusual
properties in the superconducting state. Starting from the
discovery of superconductivity in a heavy-fermion compound
CePt$_3$Si (Ref. \onlinecite{Bauer04}), the list of
noncentrosymmetric superconductors has been steadily growing and
now includes dozens of materials, such as UIr (Ref.
\onlinecite{Akazawa04}), CeRhSi$_3$ (Ref. \onlinecite{Kimura05}),
CeIrSi$_3$ (Ref. \onlinecite{Sugitani06}), Y$_2$C$_3$ (Ref.
\onlinecite{Amano04}), Li$_2$(Pd$_{1-x}$,Pt$_x$)$_3$B (Ref.
\onlinecite{LiPt-PdB}), KOs$_2$O$_6$ (Ref. \onlinecite{KOsO}), and
many others. What sets crystals without inversion symmetry apart
from their centrosymmetric counterparts (which are the ones
usually considered in the literature) is the role of the
spin-orbit (SO) coupling of electrons with the lattice potential.
In contrast to the centrosymmetric case, it qualitatively changes
the nature of the electron wavefunctions, lifting the spin
degeneracy of the Bloch bands almost everywhere in the Brillouin
zone, and resulting in a complex spin texture of the bands in the
momentum space. This has profound consequences for
superconductivity, including unusual nonuniform (``helical'')
superconducting phases,\cite{MinSam94,Agter03,DF03,Sam04,MS08},
magnetoelectric effect,\cite{Lev85,Edel89,Yip02,Fuji05} and a
strongly anisotropic spin susceptibility with a large residual
component.\cite{Edel89,GR01,FAKS04,Sam05,Sam07}

One subject that has received little attention in the recent
studies of noncentrosymmetric metals is the effects of the spin
texture and the wavefunction topology in the \textit{normal}
state. Topological properties of wavefunctions are known to play a
crucial role in many condensed matter systems. A classic example
is the integer quantum Hall effect, which was explained in Ref.
\onlinecite{TKKN82} in terms of the Chern numbers of the magnetic
Bloch bands, see also Ref. \onlinecite{Kohm85}. Other examples
include the spontaneous (anomalous) Hall effect in ferromagnetic
metals and semiconductors, whose relation to the wavefunction topology was emphasized
in Refs. \onlinecite{ON02,JNM02,Hald04}, the spin Hall
effect,\cite{Spin-Hall-1,Spin-Hall-2} the quantum spin Hall effect
in topological band insulators,\cite{KM05,BZ06,FKM07} and electric
polarization of crystalline insulators.\cite{polar-dielec}

One common feature shared by these systems is the importance of
the Berry phase of band electrons. Discovered originally in the
context of quantum systems with adiabatically changing
parameters,\cite{Berry84} the Berry phase found its applications
in many areas of physics, see, e.g., Ref.
\onlinecite{Berry-review}, and was introduced into the dynamics of
electrons in solids in Ref. \onlinecite{Zak89}. For Bloch
electrons, the role of the parameter space is played by the
reciprocal (or momentum) space: If, when subjected to a slowly
varying external field, the quasiparticle wave vector evolves
semiclassically along a closed path in the Brillouin zone, then
the wavefunction picks up a path-dependent Berry phase. It is the
Berry phase, or, more precisely, the flux of the associated Berry
curvature through the magnetic Brillouin zone in two dimensions,
that determines the quantized Hall conductivity.\cite{TKKN82}
Nonzero Berry curvature also determines the lowest-order quantum
corrections to the semiclassical dynamics of quasiparticles,
leading, in particular, to anomalous velocity terms in the
equations of motions. While the importance of such terms was
noticed several decades ago,\cite{KL54} their relation to the
topological characteristics of the band wavefunctions was
established only recently, see Refs. \onlinecite{CN95,SN99}. 

Among the effects of the anomalous velocity that are of particular
relevance to metals is the modification of the Lifshitz-Onsager relation, which describes 
the semiclassical quantization of the electron energy levels in an applied magnetic field: $A(E)=(2\pi\hbar eB/c)(n+\gamma)$ (Ref.
\onlinecite{LL9}). Here $B$ is the magnetic field, $A$ is the
cross-sectional area of a closed classical orbit of an electron in
momentum space, and $n$ is a large positive integer. As for the
correction $\gamma$, while it is common to use $\gamma=1/2$, it is 
no longer the case if the electron spin and the SO coupling are 
taken into account. This was analyzed in detail for a
centrosymmetric metal in Ref. \onlinecite{Roth66}, and revisited
recently in Refs. \onlinecite{MS98,MS99}, where it was shown that
$\gamma$ has a non-universal value, which depends on the details
of the orbit.

The goal of this paper is twofold. First, we want to present a
systematic analysis of the SO coupling of band electrons in
noncentrosymmetric crystals, which is done in Sec. \ref{sec: band
structure}. Sec. \ref{sec: SOC symmetry} focuses on the role of
point-group and time-reversal symmetries in determining the
structure of the SO coupling in the reciprocal space. In Sec.
\ref{sec: pseudospin}, we define the basis of Bloch pseudospin
eigenstates and discuss the distinction between the symmetric and
antisymmetric SO coupling, the latter being nonzero only in the
noncentrosymmetric case. In Sec. \ref{sec: one-band}, we introduce
the generalized Rashba model and provide explicit expressions for the
antisymmetric SO coupling for all noncentrosymmetric crystal 
symmetries. A distinctive feature of
noncentrosymmetric crystals is that the antisymmetric SO coupling
always vanishes, for symmetry reasons, at some isolated points or
even whole lines in the Brillouin zone, which leads to the
presence of mandatory band degeneracies, see Sec. \ref{sec: band
degen}.

Our second goal is to derive the semiclassical
equations of motion of fermionic quasiparticles in
noncentrosymmetric metals, which is done in Sec. \ref{sec: qc
dynamics}. The spin effects play a crucial role and need to be
fully taken into account: In addition to the interaction with an external
magnetic field, we include the antisymmetric SO coupling of
electron with the crystal lattice potential, as well the effects
of the exchange field in a magnetic crystal. We consider only the case of noninteracting electrons. We assume that
the band splitting caused by the SO coupling is sufficiently
strong to make it possible to treat the quasiparticle dynamics in
different nondegenerate bands independently. The semiclassical
equations of motion can be derived using various techniques, see,
e.g., the wave-packet Lagrangian formalism of Ref.
\onlinecite{SN99}. Our derivation in Sec. \ref{sec: QC eqs
derivation}, which is based on the general semiclassical analysis
for multicomponent wavefunctions developed by Littlejohn and Flynn,\cite{LF91} yields the semiclassical Hamiltonian and the
equations of motions containing terms of the zeroth as well as the first order
in Planck's constant $\hbar$. The latter can be dubbed ``quantum
corrections'', and are shown to depend on the Berry curvature of
the SO split bands. In Sec. \ref{sec: top defects}, the
contributions of the band degeneracy points and lines to the Berry
curvature are discussed. 

Finally, in Sec. \ref{sec: applications},
we apply the theory developed in the preceding sections to the
Lifshitz-Onsager quantization and the de Haas-van Alphen (dHvA)
effect in noncentrosymmetric metals (Sec. \ref{sec: dHvA}), and
also to the anomalous Hall effect (AHE) in ferromagnetic noncentrosymmetric metals (Sec. \ref{sec: AHE}).

\section{Spin-orbit coupling in noncentrosymmetric crystals}
\label{sec: band structure}

Our starting point is the following Hamiltonian for non-interacting electrons in a crystal:
\begin{equation}
\label{general H}
    \hat H=\frac{\hbp^2}{2m}+U(\br)
    +\frac{\hbar}{4m^2c^2}\hat{\bm{\sigma}}[\bm{\nabla}U(\br)\times\hbp],
\end{equation}
where $\hbp=-i\hbar\bm{\nabla}$ is the momentum operator, $U(\br)$
is the crystal lattice potential, and $\hat{\bm{\sigma}}$ are the
Pauli matrices. We neglect impurities, lattice defects, and
phonons, so that $U(\br)$ has the perfect periodicity of a Bravais
lattice. In the absence of the SO coupling, which is described by
the last term, $\hat H_{SO}$, the eigenstates of the Hamiltonian
are the Bloch spinors, labelled by the wave vector $\bk$ (which
takes values in the first Brillouin zone), the band index $n$, and
the spin index $s$:
\begin{equation}
\label{Bloch spinors}
    \langle\br\sigma|\bk n s\rangle=\frac{1}{\sqrt{\cal V}}
    \varphi_{\bk n}(\br)e^{i\bk\br}\chi_s(\sigma).
\end{equation}
Here ${\cal V}$ is the system volume, $\sigma=\uparrow,\downarrow$
is the spin projection, $\varphi_{\bk n}(\br)$ have the same
periodicity as the crystal lattice, and $\chi_s(\sigma)$ are the
basis spinors: $\chi_s(\sigma)=\delta_{s\sigma}$. The
corresponding eigenvalues have the following symmetry properties:
$\epsilon_n(\bk)=\epsilon_n(-\bk)$,
$\epsilon_n(\bk)=\epsilon_n(g^{-1}\bk)$, where $g$ is an
operation from the point group of the crystal. We will call the
electron bands calculated without the SO coupling the ``orbital''
bands.

Let us calculate the matrix elements of the electron-lattice SO
coupling in the basis of the Bloch states (\ref{Bloch spinors}):
\begin{equation}
\label{H SO matrix}
    \langle{\bk n}s|\hat H_{SO}|\bk' n's'\rangle=
    \frac{\hbar}{4m^2c^2}\sum_{ijl}e_{ijl}\langle s|\hat\sigma_i|s'\rangle\frac{1}{\cal V}\int
    d^3\br\,\Theta_{jl}(\br)e^{i(\bk'-\bk)\br},
\end{equation}
where $i,j,l=x,y,z$, and
$$
    \Theta_{jl}(\br)=(\nabla_jU)\varphi^*_{{\bk n}}(\br)(\hat p_l+\hbar k'_l)\varphi_{\bk'n'}(\br).
$$
Since $\Theta_{jl}$ are lattice-periodic functions of $\br$, the
integral in Eq. (\ref{H SO matrix}) is nonzero only if
$\bk'-\bk=\bm{G}$, where $\bm{G}$ is a reciprocal lattice vector.
Because both $\bk$ and $\bk'$ are in the first Brillouin zone, the
only possibility is $\bk'=\bk$. The Hamiltonian remains
nondiagonal in both the band and spin indices, and can be written
in the second-quantized form as follows:
\begin{equation}
\label{H gen}
    H=\sum_{\bk}\sum_{n,n'}\sum_{s,s'}[\epsilon_n(\bk)\delta_{nn'}\delta_{ss'}+
    \bm{L}_{nn'}(\bk)\bm{\sigma}_{ss'}]a^\dagger_{\bk ns}a_{\bk n's'},
\end{equation}
where $a^\dagger$ and $a$ are the electron creation and
annihilation operators, the chemical potential is included in the
band dispersion functions (we neglect the difference between the
chemical potential and the Fermi energy $\epsilon_F$), and the
functions
\begin{equation}
\label{gamma gen}
    \bm{L}_{nn'}(\bk)=\frac{\hbar}{4m^2c^2}\frac{1}{\upsilon}\int_\upsilon
    d^3\br\,\bm{\nabla}U(\br)\times\left[\varphi^*_{{\bk n}}(\br)(\hbp+\hbar\bk)\varphi_{\bk n'}(\br)\right]
\end{equation}
describe the SO coupling. The components of $\bm{L}_{nn'}$ with
$n=n'$ and $n\neq n'$ can be interpreted, respectively, as the
intraband and interband matrix elements of the orbital angular
momentum of band electrons. The integration in Eq. (\ref{gamma
gen}) is performed over the unit cell of volume $\upsilon$.  The
Hamiltonian (\ref{H gen}) is exact for non-interacting electrons,
regardless of the band structure and the strength of the SO
coupling.

\subsection{Symmetry of the SO coupling}
\label{sec: SOC symmetry}

Although one can, in principle, calculate the SO interaction
functions $\bm{L}_{nn'}(\bk)$ using Eq. (\ref{gamma gen}), it is
more convenient to treat them as parameters of the model, which
satisfy certain symmetry-imposed conditions. Since the Hamiltonian
(\ref{H gen}) is Hermitian, we have
\begin{equation}
\label{gamma herm}
    \bm{L}_{nn'}(\bk)=\bm{L}^*_{n'n}(\bk).
\end{equation}
As for the point group operations, it is sufficient to consider
the transformations under proper rotations and inversion, since
any improper operation (e.g. a mirror reflection in a plane) can
be represented as the product of a proper rotation and the
inversion $I\br=-\br$. Under a rotation $g=R$ about a direction
$\bm{n}$ by an angle $\theta$, the second-quantization operators
transform as follows: $a^\dagger_{{\bk n}s}\to\sum_{s'}
a^\dagger_{R\bk,ns'}{\cal U}_{s's}(R)$, where $\hat{\cal
U}(R)\equiv D^{(1/2)}(R)$ is the spinor representation of the
rotation, see Appendix \ref{app: transformation}. Under inversion
$I$, $a^\dagger_{{\bk n}s}\to a^\dagger_{-{\bk n}s}$. Using the
identity ${\cal U}(R)\hat{\bm{\sigma}}{\cal
U}^\dagger(R)=R^{-1}\hat{\bm{\sigma}}$, we obtain that
$\bm{L}_{nn'}(\bk)$ transform like pseudovectors:
\begin{eqnarray}
    &&R:\ \bm{L}_{nn'}(\bk)\to R\bm{L}_{nn'}(R^{-1}\bk),\\
\label{L-pseudovec}
    &&I:\ \bm{L}_{nn'}(\bk)\to \bm{L}_{nn'}(-\bk).
\end{eqnarray}
Finally, under time reversal $K$, $fa^\dagger_{{\bk n}s}\to
f^*\sum_{s'}(i\sigma_2)_{ss'}a^\dagger_{-{\bk n}s'}$
($f$ is an arbitrary $c$-number constant), therefore
\begin{equation}
    K:\ \bm{L}_{nn'}(\bk)\to-\bm{L}_{nn'}^*(-\bk).
\end{equation}
Since the Hamiltonian is invariant under all operations $g$ from the point group, we obtain:
\begin{equation}
\label{gamma g}
        \bm{L}_{nn'}(\bk)=g\bm{L}_{nn'}(g^{-1}\bk).
\end{equation}
In addition, if the time-reversal invariance is not broken, then
\begin{equation}
\label{gamma K}
        \bm{L}_{nn'}(\bk)=-\bm{L}^*_{nn'}(-\bk).
\end{equation}

In a centrosymmetric crystal,
$\bm{L}_{nn'}(\bk)=\bm{L}_{nn'}(-\bk)$, therefore it follows from Eq. (\ref{gamma K}) that
\begin{equation}
\label{bmL I}
    \bm{L}_{nn'}(\bk)=-\bm{L}_{nn'}^*(\bk).
\end{equation}
Using Eq. (\ref{gamma herm}) we obtain that the band-diagonal
matrix elements of the SO coupling vanish: $\bm{L}_{nn}(\bk)=0$.
Therefore one needs to include at least two orbital bands in Eq.
(\ref{H gen}):
$\bm{L}_{12}(\bk)=-\bm{L}_{21}(\bk)=i\bm{\ell}(\bk)$, where the
pseudovector $\bm{\ell}$ is real, even in $\bk$, and satisfies
$\bm{\ell}(\bk)=g\bm{\ell}(g^{-1}\bk)$. In contrast, in a
noncentrosymmetric crystal the constraint (\ref{bmL I}) is absent,
and the effects of SO coupling can be studied in a minimal model
with just one orbital band. Setting $n=0$, we have:
$\bm{L}_{00}(\bk)=\bgam(\bk)$, where the pseudovector $\bgam$ is
real, odd in $\bk$, and invariant with respect to the point group
operations: $\bgam(\bk)=g\bgam(g^{-1}\bk)$.

\subsection{Pseudospin representation}
\label{sec: pseudospin}

The simplest description of the SO coupling in a
noncentrosymmetric crystal is achieved in a single-band minimal
model mentioned in the end of the previous section. A serious
drawback of such a model is that it includes only the asymmetric
SO coupling and thus completely neglects the SO coupling of
electrons with atomic cores. The latter, which is insensitive to
the spatial arrangement of the atoms in the crystal, can be
important in compounds with heavy atoms. This problem can be
remedied if one formulates the theory of the electron-lattice SO
coupling using a ``pseudospin'' representation.

We begin by separating the inversion-symmetric and antisymmetric 
parts of the lattice potential:
$U(\br)=U_s(\br)+U_a(\br)$, where
$$
    U_s(\br)=\frac{U(\br)+U(-\br)}{2}, \quad U_a(\br)=\frac{U(\br)-U(-\br)}{2}.
$$
The Hamiltonian (\ref{general H}) can then be represented as
follows: $\hat H=\hat H_s+\hat H_a$, where
\begin{eqnarray}
\label{Hs}
        &&\hat H_s=\frac{\hbp^2}{2m}+U_s(\br)
        +\frac{\hbar}{4m^2c^2}\hat{\bm{\sigma}}[\bm{\nabla}U_s(\br)\times\hbp],\\
\label{Ha}
    &&\hat H_a=U_a(\br)+\frac{\hbar}{4m^2c^2}\hat{\bm{\sigma}}[\bm{\nabla}U_a(\br)\times\hbp].
\end{eqnarray}

Next, we diagonalize the inversion-symmetric part of the
Hamiltonian: $\hat
H_s|\bk\mu\alpha\rangle=\epsilon_\mu(\bk)|\bk\mu\alpha\rangle$.
The spectrum consists of the bands that are two-fold degenerate at
each $\bk$, because of the combined symmetry operation $KI$. The
index $\mu$ labels the bands, while $\alpha=1,2$ distinguishes two
orthonormal states within the same band (the ``pseudospin
states''), which are defined as follows: $|\bk\mu 2\rangle\equiv
KI|\bk\mu 1\rangle$. Explicitly:
\begin{equation}
\label{Bloch pseudospinors}
        |\bk\mu\alpha\rangle=\frac{1}{\sqrt{\cal V}}
    \left( \begin{array}{c} u_{\bk\mu\alpha}(\br)\\ v_{\bk\mu\alpha}(\br)\end{array}\right)e^{i\bk\br},\qquad
     u_{\bk\mu 2}(\br)=v^*_{\bk\mu 1}(-\br),\quad v_{\bk\mu 2}(\br)=-u^*_{\bk\mu 1}(-\br).
\end{equation}
Here $u_{\bk\mu\alpha}(\br)$ and $v_{\bk\mu\alpha}(\br)$ have the
same periodicity as the crystal lattice. There is still freedom in
the relative ``orientation'' of the eigenspinors at different
points in the Brillouin zone. Following Ref. \onlinecite{UR85}, we
choose the pseudospin states at each $\bk$ in such a way that they
transform under the point group operations (including inversion)
and time reversal in the same manner as the pure spin eigenstates, see Eq.
(\ref{transform-states}).
Starting from some wave vector $\bk_0$ in the fundamental domain
of the first Brillouin zone, one can use the expressions
\begin{eqnarray}
    &&g|\bk_0\mu\alpha\rangle=\sum_\beta|g\bk_0,\mu\beta\rangle D_{\beta\alpha}^{(1/2)}(g),\\
    &&K|\bk_0\mu\alpha\rangle=\sum_\beta(i\sigma_2)_{\alpha\beta}|-\bk_0,\mu\beta\rangle,
\end{eqnarray}
to define the pseudospin states at all other wave vectors belonging to the star of $\bk_0$.

We can now calculate the matrix elements of the antisymmetric part
of the Hamiltonian in the pseudospin basis (\ref{Bloch
pseudospinors}): $\langle\bk\mu\alpha|\hat H_a|\bk'\nu\beta\rangle=\delta_{\bk,\bk'}X_{\alpha\beta}^{\mu\nu}(\bk)$,
where $X_{\alpha\beta}^{\mu\nu}(-\bk)=-X_{\alpha\beta}^{\mu\nu}(\bk)$
due to the odd parity of $\hat H_a$. These matrix elements
can be expressed, quite generally, in terms of the Pauli matrices
in the pseudospin space as follows:
$X^{\mu\nu}_{\alpha\beta}=iA_{\mu\nu}\delta_{\alpha\beta}+\bm{B}_{\mu\nu}\bm{\sigma}_{\alpha\beta}$. 
[Note that it would be wrong to associate the first and the second
terms in this expression with the potential and the SO contributions in Eq. (\ref{Ha}), respectively. For instance,
$\langle\bk\mu\alpha|U_a|\bk\nu\beta\rangle$ is not proportional to $\delta_{\alpha\beta}$, in general.] The Hamiltonian of the
system in the pseudospin representation has the following form:
\begin{equation}
\label{H pseudospin}
    H=\sum_{\bk,\mu\nu}\sum_{\alpha\beta=1,2}[\epsilon_\mu(\bk)\delta_{\mu\nu}\delta_{\alpha\beta}+
    iA_{\mu\nu}(\bk)\delta_{\alpha\beta}+\bm{B}_{\mu\nu}(\bk)\bm{\sigma}_{\alpha\beta}]b^\dagger_{\bk\mu\alpha}b_{\bk\nu\beta}.
\end{equation}
Here, in contrast to Eq. (\ref{H gen}), the effects of the
inversion-antisymmetric part of the lattice potential (the last
two terms) are explicitly separated from the inversion-symmetric
part (the first term), the latter containing all the information
about the intra-atomic SO coupling.

The parameters of the Hamiltonian (\ref{H pseudospin}) must
satisfy a set of rather restrictive conditions, which are imposed by
the symmetry of the system. Taking into account the requirements
of Hermiticity and time-reversal invariance, see Sec. \ref{sec:
SOC symmetry}, we obtain that $A_{\mu\nu}(\bk)$ and
$\bm{B}_{\mu\nu}(\bk)$ are real, odd in $\bk$, and satisfy
$A_{\mu\nu}(\bk)=-A_{\nu\mu}(\bk)$ and
$\bm{B}_{\mu\nu}(\bk)=\bm{B}_{\nu\mu}(\bk)$. As for the point
group invariance, proper and improper operations have to be
considered separately, using the fact that, by construction, the pseudospin states transform in the same
way as the pure spinor states considered in Sec. \ref{sec: SOC
symmetry}. For a proper rotation $R$ we have: $A_{\mu\nu}(\bk)=A_{\mu\nu}(R^{-1}\bk)$ and
$\bm{B}_{\mu\nu}(\bk)=R\bm{B}_{\mu\nu}(R^{-1}\bk)$. For an improper operation which is a product of a rotation $\tilde
R$ and inversion $I$ we have: $A_{\mu\nu}(\bk)=-A_{\mu\nu}(\tilde
R^{-1}\bk)$ and $\bm{B}_{\mu\nu}(\bk)=-\tilde
R\bm{B}_{\mu\nu}(\tilde R^{-1}\bk)$.

The antisymmetric SO coupling can, in general, lift the pseudospin
degeneracy of the bands $\epsilon_\mu(\bk)$. Since the second term in Eq. (\ref{H
pseudospin}) is invariant under arbitrary rotations in the
pseudospin space, the degeneracy is removed only if
$\bm{B}_{\mu\nu}(\bk)\neq 0$. The bands always remain
at least two-fold degenerate at the center of the Brillouin
zone, because $\bm{B}_{\mu\nu}(\bm{0})=0$.

\subsection{One-band Hamiltonian}
\label{sec: one-band}

The electron band structure in noncentrosymmetric crystals has
some peculiar features, e.g., a nontrivial topology of the band
wavefunctions, which have a significant effect on the dynamics of
quasiparticles. We shall study those features using a model in
which just one band is kept. This can be justified if the energy
splitting of the two pseudospin states with the same band index
$\mu$ due to the antisymmetric SO coupling is much smaller than
the separation between the bands with different $\mu$. In the
one-band model, setting $\mu=\nu=0$ we have $A_{00}(\bk)=0$ and
$\bm{B}_{00}(\bk)\equiv\bgam(\bk)$, so that the Hamiltonian
(\ref{H pseudospin}) is reduced to the following form:
\begin{equation}
\label{H_Rashba}
    H=\sum\limits_{\bk}\sum_{\alpha,\beta=1,2}
    [\epsilon_0(\bk)\delta_{\alpha\beta}+\bgam(\bk)\bm{\sigma}_{\alpha\beta}]
    b^\dagger_{\bk\alpha}b_{\bk\beta}
\end{equation}
Here the band dispersion satisfies
$\epsilon_0(\bk)=\epsilon_0(-\bk)$ and
$\epsilon_0(\bk)=\epsilon_0(g^{-1}\bk)$. The antisymmetric
electron-lattice SO coupling is described by a real pseudovector
function $\bgam(\bk)$, which is odd in $\bk$. According
to Sec. \ref{sec: pseudospin}, it has the following symmetry
properties with respect to the point group operations: Under a
proper rotation $R$, $\bgam(\bk)=R\bgam(R^{-1}\bk)$, while under
an improper operation $I\tilde R$, $\bgam(\bk)=-\tilde
R\bgam(\tilde R^{-1}\bk)$. Lowest-order polynomial expressions for
$\bgam(\bk)$ for all 21 noncentrosymmetric point groups are given
in Table \ref{table gammas}. 
The SO coupling Hamiltonian of the form (\ref{H_Rashba}) is sometimes called the generalized Rashba model, after Ref. \onlinecite{Rashba60}, 
in which the particular case with $\bgam(\bk)=a(k_y\hat x-k_x\hat y)$ was considered. The original Rashba model
has been extensively used (see, e.g., Ref. \onlinecite{Wink-book}) to describe the properties of quasi-two-dimensional semiconductors
which are noncentrosymmetric due to the asymmetry of the confining potential.

When it is necessary to take into account the crystal
periodicity, the basis functions should be represented as the lattice Fourier series:
$\bgam(\bk)=\sum_n\bgam_n\sin\bk\bm{R}_n$, where summation goes
over the sites $\bm{R}_n$ of the Bravais lattice which cannot be
transformed one into another by inversion. For example, in the
case of a simple tetragonal lattice, which is realized in
CePt$_3$Si (point group $\mathbb{G}=\mathbf{C}_{4v}$, space group
$P4mm$), we have in the nearest-neighbor approximation:
\begin{equation}
\label{gamma C4v periodic}
    \bgam(\bk)=a(\hat x\sin k_yd-\hat y\sin k_xd),
\end{equation}
where $d$ is the lattice spacing in the basal plane. In order to obtain a nonzero $z$-component of the
SO coupling, one has to go beyond the nearest-neighbor
approximation.

\begin{table}
\caption{Representative expressions for the antisymmetric SO
coupling for all noncentrosymmetric point groups (using both
Schoenflies and International notations); $a_i$ and $a$ are real
constants, $b_i$ and $b$ are complex constants, and $k_\pm=k_x\pm
ik_y$. In the right column, the types of the symmetry-imposed
zeros of the SO coupling are listed. }
\begin{tabular}{|c|c|c|}
    \hline
    $\mathbb{G}$   & $\bgam(\bk)$ & zeros of $\bgam(\bk)$ \\ \hline
    $\mathbf{C}_1$ ($1$)      & $(a_1k_x+a_2k_y+a_3k_z)\hat x+(a_4k_x+a_5k_y+a_6k_z)\hat y+(a_7k_x+a_8k_y+a_9k_z)\hat z$ & point \\ \hline
    $\mathbf{C}_2$ ($2$)      & $(a_1k_x+a_2k_y)\hat x+(a_3k_x+a_4k_y)\hat y+a_5k_z\hat z$ & point \\ \hline
    $\mathbf{C}_s$ ($m$)      & $a_1k_z\hat x+a_2k_z\hat y+(a_3k_x+a_4k_y)\hat z$ & point \\ \hline
    $\mathbf{D}_2$ ($222$)      & $a_1k_x\hat x+a_2k_y\hat y+a_3k_z\hat z$ & point \\ \hline
    $\mathbf{C}_{2v}$ ($mm2$)      & $a_1k_y\hat x+a_2k_x\hat y+ia_3(k_+^2-k_-^2)k_z\hat z$ & line \\ \hline
    $\mathbf{C}_4$ ($4$)      & $(a_1k_x+a_2k_y)\hat x+(-a_2k_x+a_1k_y)\hat y+a_3k_z\hat z$ & point \\ \hline
    $\mathbf{S}_4$ ($\bar{4}$)      & $(a_1k_x+a_2k_y)\hat x+(a_2k_x-a_1k_y)\hat y+(bk_+^2+b^*k_-^2)k_z\hat z$ & line \\ \hline
    $\mathbf{D}_4$ ($422$)      & $a_1(k_x\hat x+k_y\hat y)+a_2k_z\hat z$ & point \\ \hline
    $\mathbf{C}_{4v}$ ($4mm$)      & $a_1(k_y\hat x-k_x\hat y)+ia_2(k_+^4-k_-^4)k_z\hat z$ & line \\ \hline
    $\mathbf{D}_{2d}$ ($\bar{4}2m$)      & $a_1(k_x\hat x-k_y\hat y)+a_2(k_+^2+k_-^2)k_z\hat z$ & line \\ \hline
    $\mathbf{C}_3$ ($3$)      & $(a_1k_x+a_2k_y)\hat x+(-a_2k_x+a_1k_y)\hat y+a_3k_z\hat z$ & point \\ \hline
    $\mathbf{D}_3$ ($32$)      & $a_1(k_x\hat x+k_y\hat y)+a_2k_z\hat z$ & point \\ \hline
    $\mathbf{C}_{3v}$ ($3m$)      & $a_1(k_y\hat x-k_x\hat y)+a_2(k_+^3+k_-^3)\hat z$ & line \\ \hline
    $\mathbf{C}_6$ ($6$)      & $(a_1k_x+a_2k_y)\hat x+(-a_2k_x+a_1k_y)\hat y+a_3k_z\hat z$ & point \\ \hline
    $\mathbf{C}_{3h}$ ($\bar{6}$)      & $(b_1k_+^2+b_1^*k_-^2)k_z\hat x+i(b_1k_+^2-b_1^*k_-^2)k_z\hat y+
                    (b_2k_+^3+b_2^*k_-^3)\hat z$ & line \\ \hline
    $\mathbf{D}_6$ ($622$)      & $a_1(k_x\hat x+k_y\hat y)+a_2k_z\hat z$ & point \\ \hline
    $\mathbf{C}_{6v}$ ($6mm$)      & $a_1(k_y\hat x-k_x\hat y)+ia_2(k_+^6-k_-^6)k_z\hat z$ & line \\ \hline
    $\mathbf{D}_{3h}$ ($\bar{6}m2$)      & $a_1[i(k_+^2-k_-^2)k_z\hat x-(k_+^2+k_-^2)k_z\hat y]+ia_2(k_+^3-k_-^3)\hat z$ & line \\ \hline
    $\mathbf{T}$ ($23$)     & $a(k_x\hat x+k_y\hat y+k_z\hat z)$ & point \\ \hline
    $\mathbf{O}$ ($432$)      & $a(k_x\hat x+k_y\hat y+k_z\hat z)$ & point \\ \hline
    $\mathbf{T}_d$ ($\bar{4}3m$)      & $a[k_x(k_y^2-k_z^2)\hat x+k_y(k_z^2-k_x^2)
        \hat y+k_z(k_x^2-k_y^2)\hat z]$ & 3 lines \\ \hline
\end{tabular}
\label{table gammas}
\end{table}

The Hamiltonian (\ref{H_Rashba}) can be diagonalized by a unitary
transformation
$b_{\bk\alpha}=\sum_{\lambda}u_{\alpha\lambda}(\bk)c_{\bk\lambda}$,
where $\lambda=\pm$, and
\begin{equation}
\label{u matrix}
    u_{1\lambda}=\frac{1}{\sqrt{2}}\sqrt{1+\lambda\frac{\gamma_z}{|\bgam|}},\qquad
    u_{2\lambda}=\frac{\lambda}{\sqrt{2}}\frac{\gamma_x+i\gamma_y}{\sqrt{\gamma_x^2+\gamma_y^2}}\sqrt{1-\lambda\frac{\gamma_z}{|\bgam|}},
\end{equation}
with the following result:
\begin{equation}
\label{H band}
    H=\sum_{\bk}\sum_{\lambda=\pm}\xi_\lambda(\bk)c^\dagger_{\bk\lambda}c_{\bk\lambda}.
\end{equation}
The energy of the fermionic quasiparticles in the $\lambda$th band
is given by
\begin{equation}
\label{Rashba-bands}
    \xi_\lambda(\bk)=\epsilon_0(\bk)+\lambda|\bgam(\bk)|,
\end{equation}
The bands are even in $\bk$ despite the antisymmetry of the SO
coupling, which is a manifestation of the Kramers degeneracy: The
states $|\bk\lambda\rangle$ and $|-\bk,\lambda\rangle$ are related
by time reversal and therefore have the same energy. In contrast, the pseudospin degeneracy of the electron
states at the same $\bk$ is lifted by the SO coupling $\bgam(\bk)$, which can be viewed as an
effective ``Zeeman magnetic field'' acting on the electron spins.
For an excitation with a given wave vector $\bk$, the spin
direction is either parallel ($\lambda=+$) or antiparallel
($\lambda=-$) to $\bgam(\bk)$. In the particular case of
$\bgam(\bk)\propto\bk$, the band index $\lambda$ has the meaning
of the helicity, which is the spin projection on the direction of
momentum. For brevity, we will be referring to the bands
(\ref{Rashba-bands}) as the helicity bands for arbitrary $\bgam(\bk)$. In real
noncentrosymmetric metals, the SO splitting between the helicity
bands is strongly anisotropic. Its magnitude can be characterized
by $E_{SO}=2\max_{\bk}|\bgam(\bk)|$. For instance, in CePt$_3$Si
$E_{SO}$ ranges from 50 to 200 meV (Ref. \onlinecite{SZB04}),
while in Li$_2$(Pd$_{1-x}$,Pt$_x$)$_3$B $E_{SO}$ is 30 meV in
Li$_2$Pd$_3$B, reaching 200 meV in Li$_2$Pt$_3$B (Ref.
\onlinecite{LP05}).

One can also consider a more general case of a magnetic
noncentrosymmetric crystal, e.g., MnSi (Ref. \onlinecite{MnSi-exp}). Neglecting the spatial variation of the
magnetization due to domain walls or a helical modulation,
the exchange field can be represented in the simplest case by a
constant pseudovector $\bm{h}$. Its effect on the electronic
structure can be described by the Hamiltonian (\ref{H_Rashba}), if
one replaces $\bgam(\bk)$ by the generalized ``Zeeman field''
$\bgam(\bk)+\bm{h}$. In this case, the time-reversal symmetry is
broken, and the helicity bands are no longer even in $\bk$:
$\xi_\lambda(\bk)=\epsilon_0(\bk)+\lambda|\bgam(\bk)+\bm{h}|\neq\xi_\lambda(-\bk)=\epsilon_0(\bk)+\lambda|\bgam(\bk)-\bm{h}|$.

\subsection{Band degeneracies}
\label{sec: band degen}

According to Table \ref{table gammas}, the helicity bands are
degenerate at some points in the Brillouin zone,
where the antisymmetric SO coupling vanishes. In particular, this
happens at $\bk=\bm{0}$, which is a trivial consequence of the
fact that $\bgam(\bk)$ is odd in $\bk$. For some point groups,
the bands are degenerate along whole lines: For the uniaxial groups
$\mathbf{C}_{2v}$, $\mathbf{S}_4$, $\mathbf{C}_{4v}$,
$\mathbf{C}_{3v}$, $\mathbf{C}_{3h}$, and $\mathbf{C}_{6v}$, as
well as for the dihedral groups $\mathbf{D}_{2d}$ and
$\mathbf{D}_{3h}$, the SO band splitting vanishes along the
principal axis $k_x=k_y=0$, while for the full tetrahedral group
$\mathbf{T}_{d}$, it vanishes along the three mutually
perpendicular axes $k_x=k_y=0$, $k_y=k_z=0$, and $k_z=k_x=0$.
Taking into account the periodicity of the reciprocal space, one
gets additional band degeneracies. For example, the SO coupling in
a simple tetragonal lattice vanishes along the following lines in
the first Brillouin zone: $k_x=0,\pm\pi/a$, $k_y=0,\pm\pi/a$, see
Eq. (\ref{gamma C4v periodic}). It will be shown in Sec. \ref{sec:
top defects} that the band degeneracies can be viewed as
``topological defects'' in $\bk$-space, which create a nontrivial
topological structure of the band electron wavefunctions.

It is important to note that the expressions from Table \ref{table gammas} 
are actually applicable to \textit{all} components of the
antisymmetric SO coupling $B_{\mu\nu}(\bk)$, see Sec. \ref{sec:
pseudospin}. Therefore, a symmetry-imposed zero of
$\bgam(\bk)$ will not be removed even if the interband elements of the
antisymmetric SO coupling are taken into account. This implies that, while 
the bands are nondegenerate almost everywhere in the Brillouin zone, each 
of the bands will \textit{always} touch al least one of the other bands at a high-symmetry point
or along a high-symmetry line. This is markedly different from the centrosymmetric case,
where the band degeneracies are possible in some cases,\cite{BSW-Zak} but they are not mandatory.  

In addition to the zeros imposed by symmetry, the antisymmetric SO
coupling might vanish at some $\bk$ for accidental reasons. It is
easy to show that, while the isolated point zeros cannot be
removed by a small variation in the parameters of the system (i.e.
are topologically stable), the accidental lines of zeros can be
removed and therefore are exceedingly improbable. The proof is
based on the observation that the SO coupling interaction
$\bgam(\bk)$ away from the degeneracies defines a mapping,
$\bk\to\hat\bgam(\bk)$, of $\bk$-space onto a sphere $S^2$, where
$\hat{\bgam}=\bgam/|\bgam|$. Since the second homotopy group of
the sphere is nontrivial: $\pi_2(S^2)=\mathbb{Z}$ (see, e.g., Ref.
\onlinecite{DFN85}), the accidental point zeros are topologically
stable. On the other hand, $\pi_1(S^2)=0$, and therefore the
accidental lines of zeros are not topologically stable. They
become stable in a ``two-dimensional'' limit $\gamma_z(\bk)\equiv
0$, which might be realized in a strongly-layered crystal, or for
an electron gas confined in a plane. In this case, the
$\hat\bgam(\bk)$ traces out a circle $S^1$, and the first homotopy
group becomes nontrivial: $\pi_1(S^1)=\mathbb{Z}$. Less formally,
the stability of the band degeneracy points can also be understood
using the following argument. Suppose there is an isolated zero of
$\bgam(\bk)$ at some $\bk=\bk_0$. If the parameters of the system
are changed so that $\bgam$ is replaced by $\bgam+\delta\bgam$,
then one can, in general, neglect the variation of $\delta\bgam$
near $\bk_0$. Assuming that $\delta\bgam$ is small enough, the new
SO coupling still has an isolated zero, at $\bk_0+\delta\bk$, where
$\delta\bk\sim{\cal O}(\delta\bgam)$. Applying this argument to a
magnetic noncentrosymmetric crystal, see Sec. \ref{sec: one-band},
we see that a sufficiently weak exchange field merely shifts the
band degeneracy point away from $\bk=\bm{0}$, but does not remove
it.

\section{Semiclassical dynamics of quasiparticles}
\label{sec: qc dynamics}

The standard derivation of the semiclassical
equations of motion\cite{AshMer} of the fermionic quasiparticles in the
$\lambda$th band in the presence of external magnetic and/or
electric fields is based on the assumption that the dynamics is generated by a classical band Hamiltonian ${\cal
H}_\lambda=\xi_\lambda(\bP/\hbar)-e\phi$, which is obtained from
the band dispersion (\ref{Rashba-bands}). Here $\bP=\bp+(e/c)\bA$
is the kinetic momentum, $\bp$ is the canonical momentum,
$\bA(\br)$ and $\phi(\br)$ are the vector and scalar potentials,
respectively, and $e$ is the absolute value of the electron
charge. When the classical trajectory is found, it can be used to
``re-quantize'' the system by inserting the corresponding action
integral into the Bohr-Sommerfeld condition to find the quantum
energy levels. It is this approach that has been extensively used,
in particular, in the theory of quantum magnetic oscillations in
metals.\cite{LL9} Its drawback is that it represents only the zeroth order term in the semiclassical expansion and therefore
does not take into account any of the important quantum
corrections. These include, in addition to the Berry phase effects
mentioned in the Introduction, also the interaction of the
magnetic moment of the band quasiparticles with the external
field.

In this section we develop a semiclassical theory of quasiparticle
dynamics near the Fermi energy in the model (\ref{H_Rashba}), with
the lowest-order quantum corrections all taken into account. It is
convenient to express the Hamiltonian in terms of the momentum
$\bp=\hbar\bk$, instead of the wave vector $\bk$, and introduce
the following notations for the quasiparticle energy and the SO
coupling:
$$
    \varepsilon(\bp)=\epsilon_F+\epsilon_0\left(\frac{\bp}{\hbar}\right),\qquad \bfg(\bp)=\bgam\left(\frac{\bp}{\hbar}\right),
$$
or $\bfg(\bp)=\bgam(\bp/\hbar)+\bm{h}$ in the magnetic case. Note 
that in this section we do not include the chemical potential in 
the definition of the quasiparticle energy, e.g., $\varepsilon(\bp)=\bp^2/2m^*$ 
in the effective mass approximation. The external electric and magnetic fields, $\bE$ and $\bB$, are 
assumed to vary slowly in comparison to the typical wavelength of 
the quasiparticles, which is of the order of the inverse Fermi wave 
vector $k_F^{-1}$. The effects of the fields are described\cite{LL9} 
by a Hamiltonian which can be written as an expansion in powers of $\bB$, 
and in which the classical canonical momentum $\bp$ is replaced by the kinetic momentum 
operator $\hat\bP=\hbp+(e/c)\bA(\br)$ (the Peierls substitution). 
In our case, this Hamiltonian has the following form:
\begin{equation}
\label{H_Rashba_fields}
    \hat H_{\alpha\beta}
    =\varepsilon(\hat\bP)\delta_{\alpha\beta}+\bfg(\hat\bP)\bm{\sigma}_{\alpha\beta}-e\phi(\br)\delta_{\alpha\beta}
	+\mu_m\bB\bm{\sigma}_{\alpha\beta}.
\end{equation}
The last term represents the interaction of the magnetic moment
of the band electrons (which contains both spin and orbital
contributions) with the magnetic field. While in general it can be
momentum-dependent and have a tensor structure, we neglect such
complications here and assume that $\mu_m=(g/2)\mu_B$, where $g$ is
the Land\'e factor and $\mu_B$ is the Bohr magneton. Note that the order of
application of the components of the kinetic momentum operator is
important, because they do not commute: $[\hat P_i,\hat
P_j]=-i(\hbar e/c)F_{ij}$, where
$F_{ij}=\nabla_iA_j-\nabla_jA_i=\sum_ke_{ijk}B_k$ is the magnetic
field written as an antisymmetric tensor. Expressions of the form
$f(\hat{\bP})$ should be understood in the following sense: If one
defines the Fourier-transform of the function $f(\bp)$ as follows:
$f(\bp)=\int d\bro\tilde f(\bro) e^{-(i/\hbar)\bro\bp}$, then
\begin{equation}
\label{P ordering}
    f(\hat\bP)=\int d\bro\,\tilde f(\bro)e^{-(i/\hbar)\bro\hat\bP}.
\end{equation}

We shall study the properties of the Hamiltonian
(\ref{H_Rashba_fields}) in the semiclassical approximation, which
is applicable when the action, $S$, calculated along a classical
trajectory is much greater than Planck's constant $\hbar$. For an
electron moving in a cyclotron orbit or radius $r_c$, $S\sim
p_Fr_c\sim\epsilon_F/\omega_c$, where $p_F=\hbar k_F$ is the
characteristic Fermi momentum and $\omega_c$ the characteristic
cyclotron frequency. The semiclassical parameter is therefore
given by
\begin{equation}
\label{qc parameter}
    \frac{\hbar\omega_c}{\epsilon_F}\ll 1.
\end{equation}
Another requirement is that the semiclassical dynamics of the
quasiparticles can be studied independently in each of the
helicity bands, i.e., in other words, there is no interband
transitions. To satisfy this requirement, one has to assume that
the spin evolves adiabatically along the classical trajectory in
such a way that the helicity is conserved.\cite{Wink-book} In the
presence of external magnetic field, the direction of the ``Zeeman
field'' $\bfg(\bp)$ is continuously changing as the excitation is
moving along its cyclotron orbit. In order for the spin to be able
to follow the direction of $\bfg(\bp)$ adiabatically, the
frequency of the orbital motion, i.e. $\omega_c$, must be much
smaller than the frequency associated with the $\bfg(\bp)$, whose
typical value is $E_{SO}/\hbar$ (assuming that the SO band
splitting remains nonzero and does not vary considerably along the
orbit, and that the exchange field in the magnetic case is
sufficiently small). Thus we have
\begin{equation}
\label{qc parameter 3}
    \frac{\hbar\omega_c}{E_{SO}}\ll 1.
\end{equation}
Finally, there are additional constraints on the magnitudes of the
applied fields, preventing the electric and magnetic breakdowns
due to the interband transitions. According to Ref.
\onlinecite{AshMer}, the following conditions should hold in order
for a single-band semiclassical description to work: $eEa\ll
E_{gap}^2/\epsilon_F$ and $\hbar\omega_c\ll E_{gap}^2/\epsilon_F$,
where the length $a$ is of the order of the lattice constant and
$E_{gap}$ is the energy splitting between the nearest bands. For
the model (\ref{H_Rashba_fields}) $E_{gap}\sim E_{SO}$, and one
can write
\begin{equation}
\label{qc parameter 2}
    \frac{\hbar\omega_c}{E_{SO}}\ll\frac{E_{SO}}{\epsilon_F}.
\end{equation}
Since typically $E_{SO}\lesssim\epsilon_F$ in noncentrosymmetric
metals, this last condition is in fact stronger than either of the
inequalities (\ref{qc parameter}) and (\ref{qc parameter 3}).

The assumption of the absence of the interband transitions fails for
sufficiently strong fields, or in the vicinity of the band
degeneracy points on the Fermi surface. We consider only those
situations, determined by the field direction and the band
structure, in which the semiclassical trajectories of the
quasiparticles in the momentum space never come close to the band
degeneracies.

\subsection{Derivation of semiclassical equations of motion}
\label{sec: QC eqs derivation}

Our derivation of the semiclassical equations of motion follows
the general scheme for multicomponent wavefunctions developed in
Ref. \onlinecite{LF91}, with some modifications relevant
specifically for a noncentrosymmetric crystal in a magnetic field, taking into account the SO coupling
and the Zeeman interaction. The idea is to
make the Hamiltonian (\ref{H_Rashba_fields}) a diagonal $2\times
2$ operator, whose matrix elements (the ``band Hamiltonians'')
would then be amenable to the usual semiclassical treatment.

We are looking for a unitary operator $\hat U(\br,\hat\bp)$, which
satisfies $\hat U^\dagger(\br,\hat\bp)\hat H(\br,\hat\bp)\hat
U(\br,\hat\bp)=\hat{\cal H}(\br,\hat\bp)$ and $\hat
U^\dagger(\br,\hat\bp)\hat U(\br,\hat\bp)=\hat\sigma_0$, where
$\hat{\cal H}=\mathrm{diag}(\hat{\cal H}_+,\hat{\cal H}_-)$. The
operators $\hat U$ and $\hat{\cal H}$ can be obtained by applying
the Weyl quantization to the corresponding Weyl symbols in a phase
space, $\hat U_W(\Gamma)$ and $\hat{\cal H}_W(\Gamma)$, where
$\Gamma=(\br,\bp)$ is a shorthand notation for canonical positions
and momenta. We recall (see, e.g., Ref. \onlinecite{ZFC-review}
and the references therein) that the Weyl symbol of an operator
$\hat A=A(\br,\hat\bp)$ is defined by the Wigner transformation as
follows:
\begin{equation}
\label{Wigner}
    A_W(\Gamma)=\int d^3\bro\;\left\langle\br+\frac{\bro}{2}\left|\hat A\right|\br-\frac{\bro}{2}\right\rangle e^{-(i/\hbar)\bro\bp}.
\end{equation}
Given the Weyl symbol, the operator is restored using
\begin{equation}
\label{Weyl}
    \langle\br|\hat A|\br'\rangle=\int\frac{d^3\bp}{(2\pi\hbar)^3}\;A_W\left(\frac{\br+\br'}{2},\bp\right) e^{(i/\hbar)\bp(\br-\br')}.
\end{equation}
The Weyl symbol of a product can be conveniently expressed by the
Moyal formula:
\begin{eqnarray}
\label{Moyal}
    (\hat A\hat B)_W(\Gamma)&=&\exp\left[\frac{i\hbar}{2}\left(\frac{\partial}{\partial\br}\frac{\partial}{\partial\bp'}-
    \frac{\partial}{\partial\bp}\frac{\partial}{\partial\br'}\right)\right]A_W(\Gamma)B_W(\Gamma')|_{\Gamma'=\Gamma}\nonumber\\
    &=&A_W(\Gamma)B_W(\Gamma)+\frac{i\hbar}{2}\left\{A_W(\Gamma),B_W(\Gamma)\right\}+{\cal O}(\hbar^2),
\end{eqnarray}
where $\{A_W,B_W\}$ is the usual Poisson bracket. In our case, the Weyl
symbols $\hat U_W(\Gamma)$ and $\hat{\cal
H}_W(\Gamma)=\mathrm{diag}[{\cal H}_+(\Gamma),{\cal H}_-(\Gamma)]$
are $2\times 2$ matrix functions in the phase space, which are
found from the equations
\begin{equation}
\label{UL_W}
    (\hat U^\dagger\hat H\hat U)_W=\hat{\cal H}_W,\quad (\hat U^\dagger\hat U)_W=\hat\sigma_0.
\end{equation}
The solutions can be sought in the form of semiclassical
expansions: $\hat U_W=\hat U_0+\hat U_1+{\cal O}(\hbar^2)$ and
${\cal H}_\lambda={\cal H}_{0,\lambda}+{\cal H}_{1,\lambda}+{\cal
O}(\hbar^2)$, where $\hat U_1$ and ${\cal H}_{1,\lambda}$ are of
the order of $\hbar$. We shall focus on the lowest two orders in
$\hbar$.

First, let us show that the semiclassical expansion of the Weyl
symbol of the Hamiltonian $\hat H$ has the following form: $\hat
H_W=\hat H_0+\hat H_1+{\cal O}(\hbar^2)$, where
\begin{equation}
\label{H0 rp}
    \hat H_0(\Gamma)=\varepsilon(\bP)\hat\sigma_0+\bfg(\bP)\hat{\bm{\sigma}}-e\phi(\br)\hat\sigma_0
\end{equation}
is the classical counterpart of Eq. (\ref{H_Rashba_fields}),
$\bP=\bp+(e/c)\bA(\br)$, and 
\begin{equation}
\label{H1 def}
	\hat H_1(\Gamma)=\mu_m\bB(\br)\hat{\bm{\sigma}}
\end{equation}
(we consider the general case, in which both the magnetic and electric fields can be nonuniform).
It is easy to see that there is no linear in $\hbar$ contributions
from the first three terms in Eq. (\ref{H_Rashba_fields}). One has
to prove this statement only for the first and second terms, which
depend on $\hat\bP$. Let us consider an operator $\hat
f=f(\hat\bP)$. Expanding the exponential in Eq. (\ref{P
ordering}), we obtain for its Weyl symbol:
\begin{equation}
\label{fWA}
    f_W=\sum_{n=0}^\infty\sum_{i_1...i_n} A_{i_1i_2...i_n}\bigl(\hat P_{i_1}\hat P_{i_2}...\hat P_{i_n}\bigr)_W,
\end{equation}
where $A_{i_1...i_n}=(1/n!)\partial^nf(\bp)/\partial
p_{i_1}...\partial p_{i_n}|_{\bp=0}$ is completely symmetric with
respect to the permutations of $i_1,i_2,...i_n$. The Weyl symbols
of the operator products on the right-hand side can be obtained by
applying the Moyal formula (\ref{Moyal}) and using the fact that
$(\hat\bP)_W=\bP$:
\begin{eqnarray*}
    &&\bigl(\hat P_{i_1}\hat P_{i_2}...\hat P_{i_n}\bigr)_W=P_{i_1}\bigl(\hat P_{i_2}\hat P_{i_3}...\hat P_{i_n}\bigr)_W
    +\frac{i\hbar}{2}\bigl\{P_{i_1},\bigl(\hat P_{i_2}\hat P_{i_3}...\hat P_{i_n}\bigr)_W\bigr\}+{\cal O}(\hbar^2)\\
    &&\quad=P_{i_1}\bigl(\hat P_{i_2}\hat P_{i_3}...\hat P_{i_n}\bigr)_W
    +\frac{i\hbar}{2}\bigl\{P_{i_1},P_{i_2}P_{i_3}...P_{i_n}\bigr\}+{\cal O}(\hbar^2)\\
    &&\quad=P_{i_1}\bigl(\hat P_{i_2}\hat P_{i_3}...\hat P_{i_n}\bigr)_W
    +\frac{i\hbar}{2}\left[\bigl\{P_{i_1},P_{i_2}\bigr\}P_{i_3}...P_{i_n}+P_{i_2}\bigl\{P_{i_1},P_{i_3}\bigr\}P_{i_4}...P_{i_n}
    +...\right]+{\cal O}(\hbar^2).
\end{eqnarray*}
The last line follows from the chain-rule property of the Poisson
bracket: $\{a,bc\}=\{a,b\}c+b\{a,c\}$. Since
$\{P_{i_1},P_{i_2}\}=-\{P_{i_2},P_{i_1}\}$, \textit{etc}, each
term in the square brackets vanishes when multiplied by the
symmetric coefficients $A_{i_1...i_n}$ and summed over
$i_1,...,i_n$. Repeating this argument for the Weyl symbols of the
remaining operator products, we find: $(\hat P_{i_1}\hat
P_{i_2}...\hat P_{i_n})_W=P_{i_1}P_{i_2}...P_{i_n}+{\cal
O}(\hbar^2)$. Substitution into Eq. (\ref{fWA}) gives
$f_W=f(\bP)+{\cal O}(\hbar^2)$. Applying this result to the matrix
elements of first two terms in the Hamiltonian
(\ref{H_Rashba_fields}), we arrive at Eq. (\ref{H0 rp}). The only
linear in $\hbar$ correction to the Weyl symbol of the Hamiltonian
comes from the magnetic moment interaction, because $\mu_m\sim\mu_B$
is proportional to Planck's constant.

In this context, one might wonder about the legitimacy of
including the SO coupling term in the classical Weyl symbol $\hat
H_0$, since formally it is also proportional to $\hbar$, which can
be traced back to the general expression for the SO interaction in
Eq. (\ref{general H}). We recall that the actual semiclassical
expansion parameter is not Planck's constant itself, but the
dimensionless ratio $\hbar\omega_c/\epsilon_F$, see Eq. (\ref{qc
parameter}). While the magnetic moment interaction is indeed
smaller than the quasiparticle energy $\varepsilon(\bp)$ by a
factor of $\mu_mB/\epsilon_F\sim\hbar\omega_c/\epsilon_F\ll 1$, the
SO coupling does not contain this ratio at all. In fact, the
latter is much greater than the former: $E_{SO}/\mu_m B\sim
E_{SO}/\hbar\omega_c\gg 1$, see Eq. (\ref{qc parameter 3}), which
justifies its treatment as a part of $\hat H_0$.

Using the Moyal formula, one can solve Eqs. (\ref{UL_W}) and
obtain the Weyl symbols of the band Hamiltonian operators: ${\cal
H}_\lambda={\cal H}_{0,\lambda}+{\cal H}_{1,\lambda}+{\cal
O}(\hbar^2)$, where $\lambda=\pm$. The leading terms in the
semiclassical expansions,
\begin{equation}
\label{Lambda0}
    {\cal H}_{0,\lambda}(\Gamma)=\varepsilon(\bP)+\lambda\left|\bfg(\bP)\right|-e\phi(\br),
\end{equation}
are just the eigenvalues of $\hat H_0$, see Eq. (\ref{H0 rp}). The
corresponding eigenvectors are given by the following expressions:
\begin{equation}
\label{taus}
    \tau_{\lambda,1}(\bP)=\frac{1}{\sqrt{2}}\sqrt{1+\lambda\frac{g_z}{|\bfg|}},\quad
    \tau_{\lambda,2}(\bP)=\frac{\lambda}{\sqrt{2}}\frac{g_x+ig_y}{\sqrt{g_x^2+g_y^2}}\sqrt{1-\lambda\frac{g_z}{|\bfg|}},
\end{equation}
cf. Eq. (\ref{u matrix}). The first quantum correction can be
represented as follows: ${\cal H}_{1,\lambda}={\cal
H}^{(1)}_{1,\lambda}+{\cal H}^{(2)}_{1,\lambda}+{\cal
H}^{(3)}_{1,\lambda}$, where
\begin{eqnarray}
\label{Lambdas}
    &&{\cal H}^{(1)}_{1,\lambda}(\Gamma)=\sum_{\alpha\beta}\tau^*_{\lambda,\alpha}H_{1,\alpha\beta}\tau_{\lambda,\beta},\nonumber\\
    &&{\cal H}^{(2)}_{1,\lambda}(\Gamma)=-\frac{i\hbar}{2}\sum_{\alpha\beta}\left(H_{0,\alpha\beta}-{\cal H}_{0,\lambda}\delta_{\alpha\beta}\right)
        \bigl\{\tau^*_{\lambda,\alpha},\tau_{\lambda,\beta}\bigr\},\\
    &&{\cal H}^{(3)}_{1,\lambda}(\Gamma)=-i\hbar\bm{\tau}^*_\lambda\bigl\{\bm{\tau}_\lambda,{\cal H}_{0,\lambda}\bigr\}.\nonumber
\end{eqnarray}
These expressions can be derived using a straightforward
generalization of the procedure described in Ref.
\onlinecite{LF91}.

The quantum band Hamiltonians $\hat{\cal H}_\lambda$ can be
derived from ${\cal H}_\lambda(\br,\bp)$ by means of the Weyl
quantization, and the wavefunctions of the quasiparticles can be
analyzed independently in the two bands. We recall that in the
semiclassical approximation, one can seek the (time-independent)
wavefunctions in the form $\psi(\br)\sim e^{iS(\br)/\hbar}$, where
the action $S$ satisfies the Hamilton-Jacobi equation ${\cal
H}_\lambda(\br,\bm{\nabla}S)=0$ (Ref. \onlinecite{MF-book}). If
the system is integrable, then a complete integral of the
Hamilton-Jacobi equation can be found, the classical orbits lie on
tori in phase space, and the motion along the orbits is
quasiperiodic. In this case the quantum energy levels can be
obtained by imposing the Bohr-Sommerfeld quantization condition on
the action integrals calculated along independent basic contours,
$C_i$, on the torus: $\oint_{C_i}\bp\,d\br=2\pi\hbar(n_i+\gamma_i)$, where $n_i$ is a large integer, and $\gamma_i$ is either $0$ or $1/2$ 
(for electrons in metals, the latter possibility is realized, see Ref.
\onlinecite{LL9}). When implementing this procedure for our
system, we encounter the problem that the classical band
Hamiltonians ${\cal H}_\lambda$ and the associated Hamilton-Jacobi
equations are not invariant under the $U(1)$ phase rotations of the
eigenvectors, $\bm{\tau}_\lambda(\Gamma)\to
e^{i\theta_\lambda(\Gamma)}\bm{\tau}_\lambda(\Gamma)$, where
$\theta_\lambda$ are arbitrary smooth functions of the phase space
coordinates. Indeed, under such transformations, ${\cal
H}_{0,\lambda}$, ${\cal H}^{(1)}_{1,\lambda}$, and ${\cal
H}^{(2)}_{1,\lambda}$ all remain the same,
but ${\cal H}^{(3)}_{1,\lambda}$ changes, since
$-i\bm{\tau}_\lambda^*\{\bm{\tau}_\lambda,{\cal
H}_{0,\lambda}\}\to-i\bm{\tau}_\lambda^*\{\bm{\tau}_\lambda,{\cal
H}_{0,\lambda}\}+\{\theta_\lambda,{\cal H}_{0,\lambda}\}$.
Furthermore, the classical band Hamiltonians are not invariant
under a usual gauge transformation of the vector potential,
$\bA(\br)\to\bA(\br)+\bm{\nabla}f$.

In order to develop a manifestly gauge-independent description of
the classical motion, we introduce, following Ref.
\onlinecite{LF91}, new coordinates in the phase space for each of
the helicity bands: $\Gamma'=(\br',\bp')$, where
$r'_i=r_i-i\hbar\bm{\tau}_\lambda^*\{\bm{\tau}_\lambda,r_i\}$ and
$p'_i=p_i-i\hbar\bm{\tau}_\lambda^*\{\bm{\tau}_\lambda,p_i\}$. It
is straightforward to show that, in the first order in $\hbar$,
the Weyl symbols in the new coordinates take the following form:
${\cal H}_\lambda(\Gamma')={\cal H}_{0,\lambda}(\Gamma')+{\cal
H}^{(1)}_{1,\lambda}(\Gamma')+{\cal
H}^{(2)}_{1,\lambda}(\Gamma')$, which no longer depends on the
phase convention for the eigenvectors $\bm{\tau}_\lambda$. The action integral along a closed path in phase space can also be
expressed in terms of the new coordinates:
\begin{equation}
\label{action int}
    J=\oint \bp\, d\br=\oint \bp'\, d\br'+i\hbar\oint\bm{\tau}_\lambda^*d\bm{\tau}_\lambda,
\end{equation}
neglecting the terms of the order of $\hbar^2$. The last term is
associated with the Berry phase picked up by the wave function in
the course of the semiclassical evolution along the closed path in
the parameter space.\cite{Berry84}

The price one pays for the restoration of the gauge invariance is that
the new coordinates are noncanonical. In particular, the new
positions have nontrivial Poisson brackets:
\begin{equation}
\label{Poisson rr prime}
    \{r'_i,r'_j\}=i\hbar\left(\frac{\partial\bm{\tau}_\lambda^*}{\partial p'_i}\frac{\partial\bm{\tau}_\lambda}{\partial p'_j}-
        \frac{\partial\bm{\tau}_\lambda^*}{\partial p'_j}\frac{\partial\bm{\tau}_\lambda}{\partial p'_i}\right).
\end{equation}
This expression can be represented in a more compact form using the fact that the eigenvectors
(\ref{taus}) depend only on the transformed kinetic momentum
$\bP'=\bp'+(e/c)\bA(\br')$. We introduce the Berry connection (or
the ``vector potential'') in the momentum space:
\begin{equation}
\label{Berry-conn-def}
    {\cal A}_{\lambda,i}(\bP)=i\bm{\tau}_\lambda^*(\bP)\frac{\partial}{\partial P_i}\bm{\tau}_\lambda(\bP),
\end{equation}
and also the corresponding curvature tensor:
\begin{equation}
\label{Berry-curv-def}
    {\cal F}_{\lambda,ij}(\bP)=\frac{\partial{\cal A}_{\lambda,j}}{\partial P_i}-\frac{\partial{\cal A}_{\lambda,i}}{\partial P_j}
    =i\left(\frac{\partial\bm{\tau}_\lambda^*}{\partial P_i}\frac{\partial\bm{\tau}_\lambda}{\partial P_j}
        -\frac{\partial\bm{\tau}_\lambda^*}{\partial P_j}\frac{\partial\bm{\tau}_\lambda}{\partial P_i}\right),
\end{equation}
which can be expressed in terms of the Berry ``magnetic field''
$\bm{{\cal B}}_\lambda(\bP)=\nablap\times\bm{{\cal
A}}_\lambda(\bP)$ as follows: ${\cal
F}_{\lambda,ij}=\sum_ke_{ijk}{\cal B}_{\lambda,k}$. Under the
phase rotation of the eigenvectors mentioned above, the Berry
connection changes: $\bm{{\cal A}}_\lambda\to\bm{{\cal
A}}_\lambda+\nablap\theta_\lambda$, while the Berry curvature
remains invariant. We shall see in Sec. \ref{sec: top defects}
that the Berry field is sensitive to the topology of the band
eigenstates. For the Poisson bracket (\ref{Poisson rr prime}) we obtain:
$\{r'_i,r'_j\}=\hbar{\cal F}_{\lambda,ij}(\bP')$. Other Poisson
brackets, $\{r'_i,p'_j\}$ and $\{p'_i,p'_j\}$, can also be expressed in terms of the Berry curvature, but they still contain
the vector potential $\bA(\br)$ and therefore are not
gauge-invariant. To fix this, we make a second change of
coordinates, to $\Gamma''=(\br',\bP')$, which makes both the
Poisson brackets and the Hamiltonian gauge-invariant and
independent of the phase choice for the $\bm{\tau}_\lambda$s.

Dropping the primes, we finally arrive at the following picture.
The classical motion of the quasiparticles in the $\lambda$th band
is described by the phase space coordinates $\Gamma=(\br,\bP)$,
which have the following Poisson brackets:
\begin{eqnarray}
\label{Poisson rP}
    	&&\{r_i,r_j\}=\hbar{\cal F}_{\lambda,ij},\nonumber\\ 
	&&\{r_i,P_j\}=\delta_{ij}+\frac{\hbar e}{c}\sum_k{\cal F}_{\lambda,ik}F_{kj},\\
    	&&\{P_i,P_j\}=-\frac{e}{c}F_{ij}-\hbar\left(\frac{e}{c}\right)^2\sum_{kl}F_{ik}{\cal F}_{\lambda,kl}F_{lj},\nonumber
\end{eqnarray}
which are manifestly noncanonical. While the noncanonical structure
of the first term in $\{P_i,P_j\}$ is not really surprising and
has a purely classical origin, the terms containing the Berry
curvature are essentially nonclassical. These terms appeared in our
derivation as the linear in $\hbar$ corrections in the
semiclassical expansion. The origin of the Berry curvature terms
can also be understood using a different argument, which is based
on the observation that $\br$ and $\bP$ refer to a particular
helicity band and therefore are not the usual positions and
momenta. Instead, they are the classical counterparts of the
\textit{band-projected} position and momentum operators, whose
commutators in the classical limit reproduce the noncanonical
Poisson brackets (\ref{Poisson rP}), see Appendix \ref{app:
band-projected r}. We would like also to mention Ref.
\onlinecite{Bliokh05}, where it was shown how a nonzero Berry
curvature appears in the adiabatic limit (i.e. when the interband
transitions are neglected), by imposing the requirement of extended
gauge invariance in the phase space on the wave packet dynamics.

We find it convenient to express both the Berry connection and the
Berry field using the spherical angle parametrization of the SO
coupling $\bfg(\bP)$: $g_x=|\bfg|\sin\alpha\cos\beta$,
$g_y=|\bfg|\sin\alpha\sin\beta$, $g_z=|\bfg|\cos\alpha$. The
eigenvectors (\ref{taus}) then become
\begin{equation}
\label{taus angular}
    \bm{\tau}_+(\bP)=\left(\begin{array}{c}
            \cos\frac{\alpha}{2}\\ e^{i\beta}\sin\frac{\alpha}{2}
                 \end{array}\right),\qquad
    \bm{\tau}_-(\bP)=\left(\begin{array}{c}
            \sin\frac{\alpha}{2}\\ -e^{i\beta}\cos\frac{\alpha}{2}
                 \end{array}\right).
\end{equation}
Inserting these expressions in Eq. (\ref{Berry-conn-def}), we obtain:
\begin{equation}
\label{Berry A angles}
    \bm{{\cal A}}_\lambda(\bP)=-\frac{1}{2}(1-\lambda\cos\alpha)\frac{\partial\beta}{\partial\bP},
\end{equation}
and
\begin{equation}
\label{Berry B angles}
    \bm{{\cal B}}_\lambda(\bP)=-\frac{\lambda}{2}\sin\alpha\left(\frac{\partial\alpha}{\partial\bP}\times\frac{\partial\beta}{\partial\bP}\right).
\end{equation}
These expressions are valid away from the band degeneracies, where
$\alpha$ and $\beta$ are not defined (and where the semiclassical
description fails anyway). In the nonmagnetic case, when the
exchange field $\bm{h}$ is absent, we have
$\bfg(-\bP)=-\bfg(\bP)$. Therefore, the inversion $\bP\to-\bP$
corresponds to $\alpha\to\pi-\alpha$ and $\beta\to\pi+\beta$,
which means that $\bm{{\cal B}}_\lambda(-\bP)=-\bm{{\cal
B}}_\lambda(\bP)$. In contrast, if $\bm{h}\neq 0$, then neither
$\bfg(\bP)$ nor the Berry field have a definite parity under
inversion.

The gauge-invariant classical band Hamiltonians are given, in the
first order in $\hbar$, by ${\cal H}_\lambda={\cal
H}_{0,\lambda}+{\cal H}^{(1)}_{1,\lambda}+{\cal
H}^{(2)}_{1,\lambda}$. Substituting here Eq. (\ref{Lambda0}) and
using the last of the Poisson brackets (\ref{Poisson rP}) in Eq.
(\ref{Lambdas}), we obtain:
\begin{equation}
\label{classical band H}
    {\cal H}_\lambda(\br,\bP)=\varepsilon(\bP)+\lambda|\bfg(\bP)|-e\phi(\br)-\hbar\bm{m}_\lambda(\bP)\bB,
\end{equation}
where
\begin{equation}
\label{cal M}
    \bm{m}_\lambda=-\frac{\mu_m}{\hbar}\sum_{\alpha\beta}\tau_{\lambda,\alpha}^*\bm{\sigma}_{\alpha\beta}\tau_{\lambda,\beta}
        -i\frac{e}{2c}\sum_{\alpha\beta}\left(\bfg\bm{\sigma}_{\alpha\beta}-
        \lambda|\bfg|\delta_{\alpha\beta}\right)\left(\frac{\partial\tau^*_{\lambda,\alpha}}{\partial\bP}\times
        \frac{\partial\tau_{\lambda,\beta}}{\partial\bP}\right)
\end{equation}
can be interpreted as the magnetic moment of the band quasiparticles (divided by $\hbar$). Using the eigenvectors (\ref{taus}), the first
term in Eq. (\ref{cal M}) can be written as
$\bm{m}_\lambda^{(1)}=-\lambda(\mu_m/\hbar)\hat{\bfg}$, where 
$\hat{\bfg}=\bfg/|\bfg|$. The second term, which is sometimes
called the Rammal-Wilkinson contribution,\cite{SN99,PST03} takes
in our case a relatively simple form. Using the spherical angle
representation (\ref{taus angular}), we obtain after some
straightforward algebra:
$$
    \bm{m}^{(2)}_\lambda=-\frac{e}{2c}|\bfg|\sin\alpha\left(\frac{\partial\alpha}{\partial\bP}\times\frac{\partial\beta}{\partial\bP}\right).
$$
Comparing this with Eq. (\ref{Berry B angles}), one finally
arrives at the following expression:
\begin{equation}
\label{M moment}
    \bm{m}_\lambda(\bP)=-\lambda\frac{\mu_m}{\hbar}\hat{\bfg}(\bP)+\lambda\frac{e}{c}|\bfg(\bP)|\bm{{\cal B}}_\lambda(\bP).
\end{equation}
The first term is consistent with the expression for the intraband
magnetic moment discussed, e.g., in Refs. \onlinecite{Sam05,Sam07} (the
negative sign appears here because the electron charge is equal to
$-e$). The second term [which is smaller than the first one by a
factor $(\hbar e/\mu_m c)(E_{SO}/p_F^2)\sim E_{SO}/\epsilon_F$], is
entirely determined by the Berry field and therefore is sensitive
to the topology of the band eigenstates. Note that, because of the
SO coupling, it would be wrong to associate the first and the
second terms in Eq. (\ref{M moment}) with, respectively, the spin
and the orbital magnetic moments.

Now we have all necessary ingredients to derive the equations of
motion for  $\br(t)$ and $\bP(t)$. The classical dynamics of the
band quasiparticles is generated by the Hamiltonians
(\ref{classical band H}) in the usual fashion:
$\dot\Gamma=\{\Gamma,{\cal H}_\lambda(\Gamma)\}$. Using the
Poisson brackets (\ref{Poisson rP}), we obtain:
\begin{equation}
\label{qc eqs of motion}
    \frac{d\br}{dt}=\bm{v}_\lambda(\bP)-\hbar\frac{d\bP}{dt}\times\bm{{\cal B}}_\lambda(\bP),\qquad
    \frac{d\bP}{dt}=-e\bm{E}(\br)-\frac{e}{c}\frac{d\br}{dt}\times\bB(\br),
\end{equation}
where $\bm{v}_\lambda=\partial{\cal H}_\lambda/\partial\bP$. While
the second of these equations has a standard Newtonian form, with
the Lorentz force on the right-hand side, the first one contains a
nonclassical term proportional to the Berry field, which is called
the ``anomalous velocity''. The importance of the anomalous term
was recognized in the early theories of the AHE in
ferromagnets,\cite{KL54} where it was derived from the interband
matrix elements of the position operator in the presence of the SO
coupling. More recently, the anomalous term appeared in the
Lagrangian wave-packet formalism of Ref. \onlinecite{SN99}, where
its relation with the Berry curvature was also clearly
established. In our derivation, the anomalous velocity appeared as
a result of the quantum corrections in the Poisson brackets
(\ref{Poisson rP}). Although the anomalous velocity term can be
viewed as the momentum-space dual of the Lorentz force, in which
the role of the magnetic field $\bB$ is played by the Berry
curvature $\bm{{\cal B}}_\lambda$, this duality is not complete.
The reason is that, unlike the physical magnetic field, which
satisfies the Maxwell equation  $\bm{\nabla}\cdot\bB=0$ and is
therefore always source-free, the Berry field has sources -- the
Berry ``magnetic monopoles'' or ``diabolical points''\cite{Berry84,Vol87} -- at the band degeneracy points, see
Sec. \ref{sec: top defects}.

To conclude this section, we recall that our analysis of the
classical equations of motion relies on the assumption that the SO
band splitting $E_{SO}$ is much greater than the energy scales
associated with the magnetic field. In the opposite limit, i.e. if
$\mu_mB\sim\hbar\omega_c\gtrsim E_{SO}$, the argument that the SO
coupling is not small in the semiclassical parameter (\ref{qc
parameter}) and can therefore be included in $\hat H_0$ [Eq.
(\ref{H0 rp})] fails. This limit requires a completely different
approach, because the inequalities (\ref{qc parameter 3}) and
(\ref{qc parameter 2}) are violated, which means that the
classical dynamics of quasiparticles can no longer be treated
separately in each helicity band. The degeneracy, or near degeneracy, of the bands changes the wavefunction geometry:
The Berry connection has to be generalized from the $U(1)$ case, see Eq. (\ref{Berry-conn-def}), to the $SU(2)$ case, in which
both basis eigenvectors, not just their phases, can be rotated.
In this case, the helicity (or the spin) becomes a dynamical variable itself, governed by an
additional equation of motion.\cite{Wink-book,BK99,CYN05}

\subsection{Band degeneracies as topological defects in momentum space}
\label{sec: top defects}

In this section, we study the relation between the band
degeneracies and the topological features in the electronic
spectrum. Let us first consider the case of an isolated point zero
in the SO coupling, symmetry-imposed or accidental, at
$\bP=\bP_0$. The former possibility is realized in many
noncentrosymmetric crystals, see Table \ref{table gammas}, where
$\bP_0=\bm{0}$. Although both the Berry connection (\ref{Berry A
angles}) and the Berry field (\ref{Berry B angles}) are not
defined at the degeneracy point, one can calculate the Berry flux,
$\oint_S\bm{{\cal B}}_\lambda\cdot d\bm{S}$, through a closed
surface $S$ in the momentum space surrounding this point. It is
straightforward to show that Eq. (\ref{Berry B angles}) can be
written as 
\begin{equation}
\label{Berry B via hat g}
	{\cal B}_{\lambda,i}=-\frac{\lambda}{4}\sum_{jk}\sum_{abc}e_{ijk}e_{abc}\,\hat g_a
	\frac{\partial\hat g_b}{\partial P_j}\frac{\partial\hat g_c}{\partial P_k},
\end{equation}
where $\hat\bfg=\bfg/|\bfg|$. For the Berry flux we have
\begin{equation}
\label{Berry flux}
    \oint_S\bm{{\cal B}}_\lambda\cdot d\bm{S}=-2\pi\lambda Q,
\end{equation}
where
\begin{equation}
\label{deg of mapping}
    Q=\frac{1}{8\pi}\sum_{ijk}e_{ijk}\oint_S dS_i\,\hat\bfg\left(\frac{\partial\hat\bfg}{\partial P_j}\times\frac{\partial\hat\bfg}{\partial P_k}\right)
        =0,\pm 1,\pm 2,...
\end{equation}
is the degree of a mapping,\cite{DFN85} $\bP\to\hat{\bfg}(\bP)$,
of the surface $S$ (which is homotopically equivalent to a sphere
$S^2$) onto the unit sphere $S^2$ corresponding to all possible
directions of $\hat\bfg$. Thus the Berry flux through a closed
surface is a topological invariant. Using the Gauss theorem in Eq.
(\ref{Berry flux}), one can write the ``Maxwell equation'' for the
Berry field: $\nablap\bm{{\cal B}}_\lambda=4\pi
q_\lambda\delta(\bP-\bP_0)$, with the right-hand side describing a
monopole at $\bP=\bP_0$, which carries the topological charge
$q_\lambda=-\lambda Q/2$. Note that, comparing Eqs. (\ref{Berry
flux}) and (\ref{Berry flux Chern}), one can also relate $Q$ with
another topological invariant, namely the Chern number for the
$\lambda$th band: $\mathrm{Ch}_\lambda=\lambda Q$.

For example, in a nonmagnetic cubic metal with the
point group $\mathbb{G}=\mathbf{O}$, we have
$\bfg(\bP)=\gamma_0\bP$, see Table \ref{table gammas}.
From Eq. (\ref{deg of mapping}) it follows that $Q=1$ and
$q_\lambda=-\lambda/2$. In general, if there is an isolated
degeneracy point at $\bP=\bm{0}$ then the SO coupling is a linear
function of the momentum near this point:
$g_i(\bP)=\sum_ja_{ij}\bP_j$. While in the triclinic case,
$\mathbb{G}=\mathbf{C}_1$, all nine coefficients here are nonzero
and different, in the higher symmetry cases some of the
coefficients vanish. For the zero to be isolated, the determinant
of the matrix $||a_{ij}||$ must be nonzero. The degree of the
mapping $\bP\to\hat{\bfg}(\bP)$ is given by the sign of this
determinant.\cite{DFN85} Therefore, the Berry field created by the
band degeneracy point has the following form:
\begin{equation}
\label{Berry monopole gen}
    \bm{{\cal B}}_\lambda(\bP)=q_\lambda\frac{\hat\bP}{|\bP|^2},\qquad q_\lambda=-\frac{\lambda}{2}\,\sign\det||a_{ij}||.
\end{equation}

An entirely different kind of the Berry field singularities is
encountered when the SO coupling vanishes along a whole line in
the momentum space. As explained in Sec. \ref{sec: band degen},
accidental lines of zeros are not topologically stable in three
dimensions. However, the lines of zeros listed in Table \ref{table
gammas} are required by the crystal symmetry and therefore
are \textit{stable}, as long as the point group is not changed by
a variation of the system's parameters. Away from a band
degeneracy line, $\bm{{\cal B}}_\lambda(\bP)$ is nonsingular and
determined by Eq. (\ref{Berry B angles}). However, along the line
the angles $\alpha$ and $\beta$ are not defined, and the Berry
field has a singularity that originates from the term containing
$\nablap\times\nablap\beta$. The precise form of the singularity
can be found by evaluating the line integral
$\Phi_\lambda^B=\oint_C\bm{{\cal A}}_\lambda\,d\bP$ along a closed
contour $C$ around the degeneracy line. According to the
definition (\ref{Berry-conn-def}), $\Phi_\lambda^B$ is nothing but
the Berry phase associated with this contour.

In order to illustrate what is happening, let us look at a
tetragonal crystal with $\mathbb{G}=\mathbf{C}_{4v}$, where the SO
coupling can be written as
\begin{equation}
\label{g C4v}
    \bfg(\bP)=\gamma_\perp(P_y\hat x-P_x\hat y)+\gamma_\parallel P_xP_yP_z(P_x^2-P_y^2),
\end{equation}
see Table \ref{table gammas} ($\gamma_\perp$ and
$\gamma_\parallel$ are constants). The SO coupling vanishes along
the line $P_x=P_y=0$. Introducing cylindrical coordinates in the
momentum space: $P_x=P_\perp\cos\varphi$,
$P_y=P_\perp\sin\varphi$, and $P_z$, we obtain for the Berry potential: $\bm{{\cal A}}_\lambda=(-1/2+f_\lambda)\nablap\varphi$,
where
$$
    f_\lambda(P_\perp,P_z,\varphi)=\frac{\lambda}{2}\frac{\gamma_\parallel P_\perp^3P_z\sin 4\varphi}{\sqrt{16\gamma_\perp^2+\gamma_\parallel^2
        P_\perp^6P_z^2\sin^24\varphi}}.
$$
We draw an arbitrary closed contour $C$ and evaluate the Berry phase integral:
\begin{equation}
\label{Berry phase C4v}
    \Phi_\lambda^B=-\pi N+\tilde\Phi_\lambda^B(C).
\end{equation}
The first term on the right-hand side contains only the winding
number $N$ of the contour around the degeneracy line, and
therefore is topologically invariant. In contrast, the second term
is not topological, because it explicitly depends on the shape and
size of the contour: For instance, if $N=1$ and both $P_\perp$ and
$P_z$ are single-valued functions of $\varphi$, then
$\tilde\Phi_\lambda^B(C)=\int_0^{2\pi}
f_\lambda[P_\perp(\varphi),P_z(\varphi),\varphi]\,d\varphi$.
Considering a small contour around the degeneracy line, one can
set $P_\perp\to 0$, then $\tilde\Phi_\lambda^B(C)\to 0$ and the
remaining topological contribution means that the Berry field has
a $\delta$-function singularity of the form
$-\pi\delta(k_x)\delta(k_y)$.

The Berry phase becomes entirely topological in the limit of a
``two-dimensional'' SO coupling, mentioned in Sec. \ref{sec: band
degen}. In this case, $g_z(\bP)=0$, and, therefore, $\bm{{\cal
A}}_\lambda=-\nablap\beta/2$ and
\begin{equation}
\label{Berry phase 2D}
    \Phi_\lambda^B=-\frac{1}{2}\oint_C d\beta(\bP)=-\pi N_\beta,
\end{equation}
where $N_\beta$ is the winding number of the angle $\beta(\bP)$
accumulated as $\bP$ moves around the contour $C$.

To summarize, the Berry field created by a line of zeros of the SO
coupling contains both the topological and nontopological
contributions. While the latter is given by Eq. (\ref{Berry B
angles}), the former is the same as that of an infinitely-thin
``solenoid'' in the momentum space coinciding with the line of
zeros. The solenoid creates a nonzero vector potential $\bm{{\cal
A}}_\lambda(\bP)$ around it, which affects the Berry phase for
contours enclosing the line of zeros, similarly to the
Aharonov-Bohm effect. In contrast, if a band degeneracy line is
present in a \textit{centrosymmetric} crystal, then one can show
that the Berry field still has a delta-function singularity at the
line but vanishes everywhere else,\cite{Blount62,MS98} i.e. the Berry field is
entirely topological. The origin of the
difference between the two cases can be understood using the fact
that in the presence of both time reversal and inversion
symmetries the band eigenstates can be chosen real. According to
Eq. (\ref{Berry-curv-def}), this means that the Berry field is
zero, except at such $\bP$ where the eigenvectors and their
derivatives are not defined, which is exactly what happens at the
degeneracy line.

\section{Selected applications}
\label{sec: applications}

\subsection{Lifshitz-Onsager relation and de Haas-van Alphen effect}
\label{sec: dHvA}

In this section we discuss manifestations of the nontrivial band
topology in a noncentrosymmetric metal. Our main focus will be on
the electron dynamics in the presence of a uniform applied field,
in particular, on the dHvA effect. The first step is to understand
how the anomalous velocity affects the cyclotron motion of the
quasiparticles and also identify the invariant tori and contours
in phase space required for the Bohr-Sommerfeld quantization. The
starting point in the quantization procedure is the expression
(\ref{action int}) for the action integral, which can be
transformed into
\begin{equation}
\label{action int rP}
    J=\oint_C\bP\,d\br-\frac{e}{c}\oint_C\bA(\br)\,d\br+\hbar\oint_C\bm{{\cal A}}_\lambda(\bP)\,d\bP,
\end{equation}
where $C$ is a closed contour. The integrals here can be
calculated using a straightforward generalization of the textbook
argument, see, e.g., Ref. \onlinecite{LL9}.

We assume a uniform applied magnetic field $\bB$, choose the
$z$-axis along the field, and set $\bm{E}=0$. Then it follows from
Eq. (\ref{qc eqs of motion}) that there are three integrals of
motion:
\begin{equation}
\label{I123}
    I_1=P_x+\frac{eB}{c}y,\quad I_2=P_y-\frac{eB}{c}x,\quad I_3=P_z,
\end{equation}
in addition to the Hamiltonian ${\cal H}_\lambda$. Equations of
motion can be transformed into the following form:
\begin{equation}
\label{eoms dHvA}
    \frac{d\br}{dt}=\bm{V}_\lambda(\bP),\qquad \frac{d\bP}{dt}=-\frac{e}{c}[\bm{V}_\lambda(\bP)\times\bB],
\end{equation}
where
$$
    V_{\lambda,x}=\frac{v_{\lambda,x}}{1+(\hbar e/c)B{\cal B}_{\lambda,z}},\ V_{\lambda,y}=\frac{v_{\lambda,y}}{1+(\hbar e/c)B{\cal B}_{\lambda,z}},\
    V_{\lambda,z}=v_{\lambda,z}+\frac{\hbar e}{c}B({\cal B}_{\lambda,x}v_{\lambda,x}+{\cal B}_{\lambda,y}v_{\lambda,y})
$$
[the singularity of $V_{\lambda,x}$ and $V_{\lambda,y}$ at ${\cal B}_{\lambda,z}(\bP)=-c/\hbar eB$ is spurious because our results 
are only valid in the first order in $\hbar$].
It follows from Eq. (\ref{eoms dHvA}) that $\dot\bP\bm{v}_\lambda=0$, therefore $P_x$ and $P_y$ trace out an
orbit in momentum space which is defined by the intersection of
the constant-energy surface ${\cal H}_\lambda(\bP)=E$ with the
plane $P_z=I_3$.  If the orbit is closed, then $P_x$ and $P_y$ are
periodic functions of time. The corresponding real-space
coordinates can be found from Eq. (\ref{I123}), from which it
follows that $x$ and $y$ also trace out a closed orbit in the
$xy$-plane, therefore the real-space trajectory is coiled around a
cylinder parallel to the $z$-axis. 

We choose the integration contour in Eq. (\ref{action int rP}) to
coincide with the momentum space orbit, with $x$ and $y$ found
from Eq. (\ref{I123}) and $z=\mathrm{const}$. The first term on
the right-hand side of Eq. (\ref{action int rP}) is
$(2c/eB)A_\lambda$, where $A_\lambda$ the area in momentum space
enclosed by the orbit. The second term is $(eB/c)A^{xy}_\lambda$,
where $A^{xy}_\lambda=(c/eB)^2A_\lambda$ is the area of the
corresponding orbit in the $xy$ plane. The last integral is the
Berry phase $\Phi_\lambda^B$ accumulated as the particle completes
one revolution along the orbit. Collecting together all three
contributions, we obtain:
\begin{equation}
\label{J A Phi}
    J=\frac{c}{eB}A_\lambda+\hbar\Phi_\lambda^B.
\end{equation}
The Berry phase term represents a linear in $\hbar$ correction to
the action integral. Since the Hamiltonian ${\cal H}_\lambda(\bP)$
also contains a quantum correction due to the magnetic moment interaction, which is given by the last term in Eq. (\ref{classical band H}),
one must, for consistency, also expand the area of the orbit:
$A_\lambda=A_{0,\lambda}+A_{1,\lambda}+{\cal O}(\hbar^2)$. The
linear in $\hbar$ correction has the following form:
$A_{1,\lambda}=\hbar
B\oint_C(dP_\perp/v_{\lambda,\perp})m_{\lambda,z}$, where
$\bP_\perp$ denotes the components of momentum perpendicular to
$\bB$, and $v_{\lambda,\perp}=|\partial{\cal
H}_{0,\lambda}/\partial\bP_\perp|$.

In arbitrary coordinate axes, a plane perpendicular to the
magnetic field is defined by the equation $\bP\hat\bB=P_0$, where
$\hat{\bB}=\bB/B$, and $P_0$ is a constant. Then $A_{0,\lambda}$
is the area of the intersection of the plane $\bP\hat\bB=P_0$ with
the constant energy surface ${\cal H}_{0,\lambda}=E$, where ${\cal
H}_{0,\lambda}(\bP)=\varepsilon(\bP)+\lambda|\bfg(\bP)|$, see Eq.
(\ref{Lambda0}). We refer to this intersection as the classical
orbit $C_\lambda(E,P_0)$. Imposing the Bohr-Sommerfeld quantization
condition on the action integral (\ref{J A Phi}) and neglecting
the terms of the order of $\hbar^2$, we arrive at the following
equation:
\begin{equation}
\label{BZ quant}
    \tilde A_\lambda(E,P_0)=\frac{2\pi\hbar eB}{c}\left(n+\frac{1}{2}\right),
\end{equation}
which implicitly determines the quasiparticle energy levels
$E_{\lambda,n}(P_0)$ in the $\lambda$th band. The area of the
classical orbit is modified by the quantum corrections as follows:
\begin{equation}
\label{tilde A def}
    \tilde A_\lambda=A_{0,\lambda}+\frac{\hbar eB}{c}(\Phi_\lambda^m+\Phi_\lambda^B),
\end{equation}
where
\begin{equation}
\label{Phi m}
    \Phi_\lambda^m(E,P_0)=\frac{c}{e}\oint_{C_\lambda}\frac{\bm{m}_\lambda(\bP)\hat\bB}{v_{\lambda,\perp}(\bP)}\,dP_\perp,
\end{equation}
results from the deformation of the orbit by the interaction of
the magnetic moment (\ref{M moment}) with the applied field, and
\begin{equation}
\label{Phi B}
    \Phi_\lambda^B(E,P_0)=\oint_{C_\lambda}\bm{{\cal A}}_\lambda(\bP)\,d\bP,
\end{equation}
is the Berry phase associated with the orbit. Thus we have
reproduced the Lifshitz-Onsager relation, with
$\gamma_\lambda=1/2-(\Phi_\lambda^m+\Phi_\lambda^B)/2\pi$, the
deviation from the universal value $1/2$ being due to the quantum
corrections to semiclassical dynamics.

The orbital quantization of the energy levels leads to a variety
of magnetooscillation phenomena, including the dHvA effect. The
oscillatory behaviour of the magnetization as a function of the
applied field is described by the Lifshitz-Kosevich formula, which
relates the dHvA frequencies to the extremal, with respect to
$P_0$, cross-sections of the Fermi surface. In our case it follows from the quantization condition (\ref{BZ quant}) that, instead
of the usual geometrical area of the cross-section, one must use the modified area given by Eq.
(\ref{tilde A def}). Including both helicity bands, we obtain the
oscillating contribution to the magnetization along the field:
\begin{equation}
\label{LifKos}
    M=\sum_\lambda\sum_{ex} M^{ex}_{\lambda}\sin\left(\frac{2\pi F^{ex}_\lambda}{B}\pm\frac{\pi}{4}\right),
\end{equation}
where the summation goes over all the extremal cross-sections,
$M^{ex}_\lambda$ are the amplitudes of the oscillations, and the
plus and minus signs in the phase shifts correspond to minimum and
maximum cross-sectional areas, respectively, see Ref.
\onlinecite{LL9}. The frequencies of the oscillations are given by
$F^{ex}_\lambda=(c/2\pi\hbar e)\tilde A^{ex}_\lambda$, where
$\tilde A^{ex}_\lambda$ is the value of expression (\ref{tilde A
def}) at the extremum, with $E=\epsilon_F$. The extremum can be shifted away from its classical position (which corresponds to 
the extremum of $A_{0,\lambda}$) due to $\Phi_\lambda^m$ and $\Phi_\lambda^B$, but this effect can be
neglected, since it produces a correction to the area that is
quadratic in the semiclassical parameter. We note that Eq.
(\ref{LifKos}) is approximate: In addition to the fundamental
harmonics, the observed dHvA signal also contains higher harmonics
with the frequencies given by integer multiples of
$F^{ex}_\lambda$.

As a simple illustration, let us consider a nonmagnetic cubic
metal with $\mathbb{G}=\mathbf{O}$. Real-life examples of this
symmetry include the Li$_2$(Pd$_{1-x}$,Pt$_x$)$_3$B family of
materials. We assume a parabolic band,
$\varepsilon(\bp)=\bp^2/2m^*$, with the effective mass $m^*$, and
use $\bfg(\bP)=\gamma_0\bP$ for the SO coupling, see Table
\ref{table gammas}. The Fermi surfaces are spheres of radii
$P_{F,\lambda}$, and the extremal classical orbits are two great
circles perpendicular to the field, therefore
$A^{ex}_{0,\lambda}=\pi P_{F,\lambda}^2$. Since the extremum is in
fact the maximum, one must use the negative sign in Eq. (\ref{LifKos}). There is an
isolated band degeneracy at $\bP=\bm{0}$, which creates a
monopole-like Berry field $\bm{{\cal
B}}_\lambda(\bP)=-(\lambda/2P_{F,\lambda}^2)\hat\bP$, see Eq.
(\ref{Berry monopole gen}). From Eq. (\ref{M moment}) it follows
that $\bm{m}_\lambda\hat\bB=0$ at the orbit, therefore
$\Phi_\lambda^m=0$. The Berry phase is nonzero and given by
$\Phi_\lambda^B=-\lambda\pi$, which yields a field-dependent
correction to the dHvA frequencies:
$F^{ex}_\lambda=cP_{F,\lambda}^2/2\hbar e-\lambda B/2$. From Eq.
(\ref{LifKos}) we obtain:
\begin{equation}
\label{LifKos-cubic}
    M=-\sum_\lambda M_\lambda\sin\left(\frac{\pi cP_{F,\lambda}^2}{\hbar eB}-\frac{\pi}{4}\right),
\end{equation}
i.e. the phase of the magnetization oscillations is shifted by
$180^{\circ}$ compared to the Lifshitz-Kosevich result without the
quantum corrections. Such a phase shift can be measured in the dHvA experiments.\cite{LK04} For the Lifshitz-Onsager parameter, we have
$\gamma=0$ (the values $\gamma=0$ and $\gamma=1$ are equivalent).
One can expect that higher orders in the semiclassical expansion
will produce corrections of the order of $B^2$ to the dHvA
frequencies, which will give rise to a magnetic field dependence
of the phase shifts in the Lifshitz-Kosevich formula (one such
correction was discussed in Ref. \onlinecite{MS05}).

Previous works on the semiclassical electron dynamics in magnetic
field have focused almost exclusively on the centrosymmetric case,
where one can show\cite{Roth66,MS98} that Lifshitz-Onsager's
$\gamma$ also differs from $1/2$ in the general case, i.e. in the
presence of both the SO coupling and the Zeeman interaction. The
deviation is non-universal in the sense that, similar to our Eqs.
(\ref{Phi m}) and (\ref{Phi B}), it depends on the details of the
classical orbit, and can be interpreted in terms of the evolution
of a classical spin vector along the orbit. We recall that, in the
centrosymmetric case, the SO coupling does not remove the spin
degeneracy of the bands, therefore the quasiparticle equations of
motion must take into account the transitions between the states
with opposite spin projections. This is also true in a 
``weakly-noncentrosymmetric'' case, when the SO band splitting is small compared to the energy scales associated with the 
external magnetic field, see the last paragraph of Sec.
\ref{sec: QC eqs derivation}. We would like to mention also the studies 
of magnetic oscillation phenomena using the Gutzwiller trace formula, 
which must be modified in the presence of the SO coupling to include
additional factors describing the classical spin evolution.\cite{BK99,KW02}

Interestingly, $\gamma$ is not necessarily equal to $1/2$, even
when the electron spin is completely neglected, which formally
corresponds to setting $\bfg(\bp)=0$ and $\mu_m=0$. In this case,
the magnetic moment vanishes and the correction to $\gamma$ is
entirely due to the Berry phase, see Ref. \onlinecite{Zak89} and
especially Ref. \onlinecite{SN99}, where the spinless limit was
studied using the wave-packet formalism. As mentioned in the end
of Sec. \ref{sec: band degen}, the Berry phase is either $\pm\pi$
or $0$, depending on whether or not the orbit encloses a band
degeneracy line in the Brillouin zone, therefore in the spinless
case there are only two possibilities: $\gamma=0$ or $\gamma=1/2$
(Ref. \onlinecite{MS99}).

\subsection{Magnetic crystals: Anomalous Hall effect}
\label{sec: AHE}

Another possible application of the formalism developed in Sec.
\ref{sec: qc dynamics} is the AHE, which is the appearance of a
transverse component of the electric current in ferromagnetic
substances, even in the absence of external magnetic field (for
recent reviews see, e.g., Refs. \onlinecite{Sini08} and \onlinecite{NSOMO09}). Semiclassical
theory of this phenomenon can be obtained by setting $\bB=0$ in
Eq. (\ref{qc eqs of motion}), which yields:
\begin{equation}
\label{AHE r dot}
    \frac{d\br}{dt}=\bm{v}_\lambda(\bp)+\hbar e\bm{E}\times\bm{{\cal B}}_\lambda(\bp).
\end{equation}
We assume a uniform electric field and neglect the internal
induction due to magnetization, which is known to be too small to
account for the AHE. The ``intrinsic'' Hall current originates
from the second term in Eq. (\ref{AHE r dot}) and can be obtained
by adding the contributions from both helicity bands:
\begin{equation}
\label{Hall current}
    \bm{j}_H=-e^2\hbar\sum_\lambda\sum_{\bp}[\bm{E}\times\bm{{\cal B}}_\lambda(\bp)]f_\lambda(\bp),
\end{equation}
where $f_\lambda(\bp)=[e^{\beta\xi_\lambda(\bp)}+1]^{-1}$ is the
quasiparticle distribution function ($\beta=1/k_BT$), and
$\xi_\lambda(\bP)=\varepsilon(\bP)+\lambda|\bfg(\bP)|-\epsilon_F$. It follows from Eq. (\ref{Hall current}) 
that the anomalous Hall current vanishes in the absence of time-reversal symmetry breaking, i.e. at
$\bm{h}=0$, in which case $\bm{{\cal B}}_\lambda$ is odd in $\bp$
(see Sec. \ref{sec: QC eqs derivation}), while $f_\lambda$ is even.
We see that the intrinsic AHE is essentially an equilibrium phenomenon, which is
related to the Berry curvature of the band wavefunctions, see
Refs. \onlinecite{ON02,JNM02}, and \onlinecite{Naga06}. Moreover, using the expression 
$\bm{{\cal B}}_\lambda=\nabla_{\bp}\times\bm{{\cal A}}_\lambda$ and integrating Eq. (\ref{Hall current}) by parts, one
can show that $\bm{j}_H$ is determined by the quasiparticle properties near the Fermi surface. This means
that not only the semiclassical description of the AHE is legitimate, but it can also be extended to interacting systems
using the standard Fermi-liquid theory arguments.\cite{Hald04} 

In two dimensions, one obtains from Eq. (\ref{Hall current}) the Hall conductivity in the form $\sigma_{xy}=-(e^2/2\pi h)\int_{\mathrm{FBZ}}dp_xdp_y\sum_\lambda
{\cal B}_{\lambda,z}f_\lambda$, where $h=2\pi\hbar$ and the momentum integration is performed over the first Brillouin zone. 
Using Eq. (\ref{Berry B via hat g}) this can be written as
\begin{equation}
\label{sigma xy 2D}
 	\sigma_{xy}=\frac{e^2}{h}\frac{1}{4\pi}\int_{\mathrm{FBZ}} dp_xdp_y\,
	\hat\bfg\left(\frac{\partial\hat\bfg}{\partial p_x}\times\frac{\partial\hat\bfg}{\partial p_y}\right)\left[f_+(\bp)-f_-(\bp)\right].
\end{equation}
The last expression takes a particularly appealing form in the case of a magnetic \textit{insulator}, which is realized when the
chemical potential lies in the gap between the ``+'' and ``-'' bands. The band gap is given by $\min[\varepsilon(\bp)+|\bfg(\bp)|]-\max[\varepsilon(\bp)-|\bfg(\bp)|]$, which vanishes in the nonmagnetic case due to the mandatory zeros of the 
SO coupling, see Sec. \ref{sec: band degen}. At zero temperature, Eq. (\ref{sigma xy 2D}) becomes
\begin{equation}
 	\sigma_{xy}=-\frac{e^2}{h}\frac{1}{4\pi}\int_{\mathrm{FBZ}} dp_xdp_y\,
	\hat\bfg\left(\frac{\partial\hat\bfg}{\partial p_x}\times\frac{\partial\hat\bfg}{\partial p_y}\right)=-\frac{e^2}{h}Q,
\end{equation}
where $Q=0,\pm 1,...$ is the degree of a mapping of the Brillouin zone (a torus) onto a unit sphere $S^2$, cf. Eq. (\ref{deg of mapping}).
Thus we come to the conclusion that the anomalous Hall conductivity in a two-dimensional ferromagnetic noncentrosymmetric insulator is quantized, 
in agreement with the result of Ref. \onlinecite{QWZ06}, which was obtained by a Kubo formula calculation. This phenomenon is similar to the integer quantum Hall effect in an external magnetic field,\cite{TKKN82} both originating from the quantization of the Berry flux through a two-dimensional Brillouin zone. 

We would like to note that the simple semiclassical picture of the AHE apparently fails in a
perfect lattice,\cite{Smit55} where, according to the second of Eq.
(\ref{qc eqs of motion}), we have $\bp(t)=\bp(0)-e\bm{E}t$. Due to
the crystal periodicity, both $\bm{v}_\lambda$ and $\bm{{\cal
B}}_\lambda$ are periodic functions of $\bp$, which means that the
quasiparticle's velocity $\dot{\br}$ is a periodic, or at least
bounded, function of time. The quasiparticle oscillates in space
and therefore cannot carry any net electric current. However, this
argument against the intrinsic AHE is not fully satisfactory,
because the linearly increasing momentum, when mapped back into
the first Brillouin zone, can eventually pass arbitrarily close to
a band degeneracy, where the anomalous velocity is singular and
the semiclassical description does not work. In addition, the case
of a perfectly periodic crystal is rather unphysical, because of
various scattering processes always present in real materials.
More detailed discussion of the AHE in ferromagnetic noncentrosymmetric metals, which should include the scattering
effects, is beyond the scope of this paper.

\section{Conclusions}
\label{sec: conclusions}

The symmetric and antisymmetric contributions to the
electron-lattice SO coupling in crystals without inversion symmetry play qualitatively different roles.
While the former just replaces spin with pseudospin,
preserving the twofold degeneracy of the bands, the
latter removes the band degeneracy almost everywhere in the Brillouin zone and creates a nonzero Berry curvature
of the resulting helicity bands. In contrast to the centrosymmetric case, there are always
remaining band degeneracies in each band. The anisotropy of the SO coupling, in particular the type and location of the
band degeneracies, is determined by the crystal symmetry, see Table \ref{table gammas}.

Using a reduced one-band model of the antisymmetric SO coupling (the generalized Rashba model),
we derived the semiclassical equations of motion of the
quasiparticles in the helicity bands, taking into account all
effects associated with the electron spin. We have found two
distinct types of quantum corrections to the semiclassical dynamics: One, which is entirely due to the
Berry curvature of the bands, makes the Poisson brackets noncanonical and
results in the anomalous velocity term in the equations of motion.
The other, which appears directly in the classical Hamiltonian,
describes the interaction of the magnetic moment of the band
quasiparticles with the applied magnetic field. The magnetic moment contains both spin and orbital contributions mixed by
the SO coupling and is also affected by the Berry curvature. Both types of the quantum corrections modify the
Bohr-Sommerfeld quantization condition, which makes them observable, e.g., in the dHvA effect.

We have considered only the case of noninteracting electrons in a normal noncentrosymmetric metal. One can expect that the 
nontrivial topology of the band wavefunctions should also affect the properties of interacting systems.  
Although there have been some interesting recent developments, see Refs. \onlinecite{Hald04,MN03,SB06}, this subject remains largely unexplored.

\acknowledgements

The author is grateful to the organizers and participants of the Workshop ``Spin Manipulation and Spin-Orbit Coupling in Semiconductors and Superconductors'' 
in Choroni, Venezuela, where this work was started, and to R. Winkler for stimulating discussions.
This work was supported by a Discovery Grant from the Natural Sciences and Engineering Research Council of Canada.

\appendix

\section{Transformation properties of quantum states and operators}
\label{app: transformation}

We adopt the convention that a symmetry transformation, either a
point group operation $g$ or time reversal $K$, changes the
physical state of the system, not the coordinate axes. Under a
proper rotation $g=R$, the position vector $\br$ is transformed
into $\br'=R\br$, where $R\equiv D^{(1)}(R)$ is the rotation
matrix in the spin-$1$ representation. We recall that the rotation
about a direction $\bm{n}$ by an angle $\theta$ ($\theta$ is
positive for a counterclockwise rotation) in the spin-$J$
representation in described by the matrix
$D^{(J)}(R)=\exp(-i\theta\bm{n}\hat{\bm{J}})$, where
$\hat{\bm{J}}$ are the generators of rotations. Under an improper
operation $g=IR$, which is represented as a product of a rotation
$R$ and inversion $I$, we have $\br\to\br'=-R\br$. In particular,
for spin-1/2 particles, we have
$D^{(1/2)}(g)=D^{(1/2)}(R)=e^{-i\theta(\bm{n}\hat{\bm{\sigma}})/2}$
for both $g=R$ and $g=IR$, since inversion does not affect the
spin degrees of freedom.

Let us consider non-interacting electrons in an ideal crystal
lattice, see Eq. (\ref{general H}). The action of the point group
and time reversal operations on spinor wavefunctions is discussed,
e.g., in Ref. \onlinecite{Messiah}. Neglecting the SO coupling and
omitting the orbital band index, the eigenstates are the Bloch
spinors $\langle\br\sigma|\bk
s\rangle\equiv\psi_{\bk,s}(\br,\sigma)=\Phi_{\bk}(\br)\delta_{s\sigma}$,
where $\Phi_{\bk}(\br)={\cal V}^{-1/2}
\varphi_{\bk}(\br)e^{i\bk\br}$, see Eq. (\ref{Bloch spinors}).
Under a point group operation $g$, these transform into
\begin{equation}
\label{g on Bloch psi}
    g\psi_{\bk,s}(\br,\sigma)=\sum_{\sigma'}D_{\sigma\sigma'}^{(1/2)}(g)\psi_{\bk,s}(g^{-1}\br,\sigma')=D_{\sigma s}^{(1/2)}(g)\Phi_{\bk}(g^{-1}\br)
    =\sum_{s'}\psi_{g\bk,s'}(\br,\sigma)D_{s's}^{(1/2)}(g).
\end{equation}
Here we used the fact that $\Phi_{\bk}(g^{-1}\br)$ corresponds to
the wave vector $g\bk$. Under time reversal,
\begin{equation}
\label{K on Bloch psi}
    K\psi_{\bk,s}(\br,\sigma)=\sum_{\sigma'}(-i\sigma_2)_{\sigma\sigma'}\psi^*_{\bk,s}(\br,\sigma')=(-i\sigma_2)_{\sigma s}\Phi^*_{\bk}(\br)
    =\sum_{s'}(i\sigma_2)_{ss'}\psi_{-\bk,s'},
\end{equation}
because $\Phi^*_{\bk}(\br)$ corresponds to the wave vector $-\bk$.

Thus we obtain that the Bloch eigenstates transform under the
point group operations and time reversal as follows:
\begin{equation}
\label{transform-states}
    g|\bk s\rangle=\sum_{s'}|g\bk,s'\rangle D_{s's}^{(1/2)}(g),\qquad K\left(f|\bk s\rangle\right)=f^*\sum_{s'}(i\sigma_2)_{ss'}|-\bk,s'\rangle.
\end{equation}
We included a constant $f$ in the second of these expressions to
highlight the antilinearity of the time reversal operation. The
transformation rules for the second quantization operators follow
immediately from Eq. (\ref{transform-states}), if one views the
Bloch eigenstates as vectors in the Fock space: $|\bk
s\rangle=a^\dagger_{\bk s}|0\rangle$, where $|0\rangle$ is the
vacuum state, and assumes that the vacuum is invariant under all
symmetry operations.\cite{ED-Book}

\section{Band-projected position operators}
\label{app: band-projected r}

In this Appendix we discuss the physical origin of the
noncanonical Poisson brackets in the semiclassical dynamics of
quasiparticles in a given helicity band. We neglect external
fields and focus on the first of the expressions (\ref{Poisson
rP}). The Poisson brackets can be obtained in the classical limit
from the commutator of the ``band-projected'' position operators
$\hat\br_\lambda=\hat\Pi_\lambda\hat\br\hat\Pi_\lambda$, where
$\hat\br$ is the usual position operator
($\hat\br=i\bm\nabla_{\bk}$ in the $\bk$-representation), and
$\hat\Pi_\lambda(\bk)\equiv|\bk\lambda\rangle\langle\bk\lambda|$
are the operators projecting onto the $\lambda$th band. Here and
below no summation over repeated band indices is assumed. Since
$\hat\Pi_\lambda^2=\hat\Pi_\lambda$, we obtain:
\begin{equation}
    \hat\br_\lambda=i\hat\Pi_\lambda\frac{\partial}{\partial\bk}\hat\Pi_\lambda=i\hat\Pi_\lambda\frac{\partial}{\partial\bk}+\hat{\bm{\Omega}}_\lambda, \qquad
    \hat{\bm{\Omega}}_\lambda=i\hat\Pi_\lambda\frac{\partial\hat\Pi_\lambda}{\partial\bk}.
\end{equation}
The band-projected position operators are $U(1)$ gauge covariant, in the
following sense: An arbitrary phase rotation of the wavefunctions
in the reciprocal space, $\psi(\bk)\to e^{i\chi(\bk)}\psi(\bk)$,
leaves the matrix elements of $\hat\br_\lambda$ invariant, if it
is accompanied by changing
$\hat{\bm{\Omega}}_\lambda\to\hat{\bm{\Omega}}_\lambda+(\bm\nabla_{\bk}\chi)\hat\Pi_\lambda$.
This variation of $\hat{\bm{\Omega}}_\lambda$ can be achieved by
redefining the phases of the eigenstates: $|\bk\lambda\rangle\to
e^{i\theta_\lambda(\bk)}|\bk\lambda\rangle$, with
$\theta_\lambda(\bk)=-\chi(\bk)$, which does not affect
$\hat\Pi_\lambda$.

Using the fact that
$\hat\Pi_\lambda(\bm\nabla\hat\Pi_\lambda)\hat\Pi_\lambda=0$, the
commutator of the band-projected positions can be represented as
follows: $[\hat r_{\lambda,i},\hat
r_{\lambda,j}]=i\hat\Pi_\lambda(\nabla_i\hat\Omega_{\lambda,j}-\nabla_j\hat\Omega_{\lambda,i})$.
It is diagonal both in the wave vector $\bk$ and the helicity
$\lambda$, with the matrix elements given by
\begin{equation}
\label{band rr comm}
    \langle\bk\lambda|[\hat r_{\lambda,i},\hat r_{\lambda,j}]|\bk\lambda\rangle=\tr\bigl(\hat\Pi_\lambda[\hat r_{\lambda,i},\hat r_{\lambda,j}]\bigr)
        =\tr\biggl(\frac{\partial\hat\Pi_\lambda}{\partial k_i}\hat\Pi_\lambda\frac{\partial\hat\Pi_\lambda}{\partial k_j}-
        \frac{\partial\hat\Pi_\lambda}{\partial k_j}\hat\Pi_\lambda\frac{\partial\hat\Pi_\lambda}{\partial k_i}\biggr).
\end{equation}
In the spin (or pseudospin) representation, the band projection
operators have the form
$\Pi_{\lambda,\alpha\beta}(\bk)=u_{\alpha\lambda}(\bk)u^\dagger_{\beta\lambda}(\bk)$,
where the unitary matrix $\hat u(\bk)$ is given by Eq. (\ref{u
matrix}). Inserting this in Eq. (\ref{band rr comm}), we obtain:
\begin{equation}
\label{comm rr uu}
    \langle\bk\lambda|[\hat r_{\lambda,i},\hat r_{\lambda,j}]|\bk\lambda\rangle=-\sum_\alpha\left(
    \frac{\partial u^*_{\alpha\lambda}}{\partial k_i}\frac{\partial u_{\alpha\lambda}}{\partial k_j}
    -\frac{\partial u^*_{\alpha\lambda}}{\partial k_j}\frac{\partial u_{\alpha\lambda}}{\partial k_i}\right).
\end{equation}

In the classical limit, the commutator is replaced by the Poisson
bracket: $[\hat r_{\lambda,i},\hat r_{\lambda,j}]\to
i\hbar\{r_{\lambda,i},r_{\lambda,j}\}$. Expressing the derivatives
in Eq. (\ref{comm rr uu}) in terms of the canonical momentum $\bp=\hbar\bk$
and introducing the eigenvectors $\bm{\tau}_\lambda$, such that
$\tau_{\lambda,\alpha}(\bp)=u_{\alpha\lambda}(\bp/\hbar)$, we
arrive at the following expression:
\begin{equation}
\label{band rr Poisson br}
    \{r_i,r_j\}=i\hbar\left(\frac{\partial\bm{\tau}_\lambda^*}{\partial p_i}\frac{\partial\bm{\tau}_\lambda}{\partial p_j}
        -\frac{\partial\bm{\tau}_\lambda^*}{\partial p_j}\frac{\partial\bm{\tau}_\lambda}{\partial p_i}\right)=\hbar{\cal F}_{\lambda,ij}(\bp),
\end{equation}
where ${\cal F}_{\lambda,ij}$ is the Berry curvature tensor in the momentum space, 
defined by Eq. (\ref{Berry-curv-def}). Thus we have recovered the first of Eqs. (\ref{Poisson rP}).

It is instructive also to interpret our results using the language of
differential geometry. The antisymmetric tensor ${\cal
F}_{\lambda,ij}$ can be used to define the Berry curvature 2-form
in the $\lambda$th band as follows:
$\omega_{B,\lambda}=(1/2)\sum_{ij}{\cal
F}_{\lambda,ij}(\bp)dp_i\wedge dp_j$, where $dp_i\wedge
dp_j=-dp_j\wedge dp_i$ is the wedge product.\cite{DFN85} Comparing
the right-hand sides of Eqs. (\ref{band rr comm}) and (\ref{band
rr Poisson br}) and expressing the band projection operators in
terms of $\bp$, we obtain:
$$
    \omega_{B,\lambda}=i\sum_{ij}\tr\biggl(\hat\Pi_\lambda\frac{\partial\hat\Pi_\lambda}{\partial p_i}\frac{\partial\hat\Pi_\lambda}{\partial p_j}\biggr)
    dp_i\wedge dp_j=i\tr\bigl(\hat\Pi_\lambda d\hat\Pi_\lambda\wedge d\hat\Pi_\lambda\bigr).
$$
In these notations the flux of the Berry field through a closed
surface $S$ in the momentum space is given by the following
expression:
\begin{equation}
\label{Berry flux Chern}
    \oint_S\bm{{\cal B}}_\lambda\cdot d\bm{S}=\int_S\omega_{B,\lambda}=-2\pi\left[\frac{1}{2\pi i}\int_S\tr\bigl(\hat\Pi_\lambda d\hat\Pi_\lambda
    \wedge d\hat\Pi_\lambda\bigr)\right]=-2\pi\,\mathrm{Ch}_\lambda.
\end{equation}
The expression in the square brackets is an integer, which is
known as the (first) Chern number for the $\lambda$th band. The
Chern numbers are probably best known in condensed matter physics
for the role they play in explaining the integer quantum Hall
effect.\cite{TKKN82,Kohm85} In Eq. (\ref{Berry flux Chern}) we
used the expression for the Chern numbers in terms of the band
projection operators from Ref. \onlinecite{ASS83}.


\begin{thebibliography}{99}

\bibitem{Bauer04}
E. Bauer, G. Hilscher, H. Michor, Ch. Paul, E. W. Scheidt, A.
Gribanov, Yu. Seropegin, H. No\"el, M. Sigrist, and P. Rogl, Phys.
Rev. Lett. \textbf{92}, 027003 (2004).

\bibitem{Akazawa04}
T. Akazawa, H. Hidaka, T. Fujiwara, T. C. Kobayashi, E. Yamamoto,
Y. Haga, R. Settai, and Y. Onuki, J. Phys.: Condens. Matter
\textbf{16}, L29 (2004).

\bibitem{Kimura05}
N. Kimura, K. Ito, K. Saitoh, Y. Umeda, H. Aoki, T. Terashima,
Phys. Rev. Lett. \textbf{95}, 247004 (2005).

\bibitem{Sugitani06}
I. Sugitani, Y. Okuda, H. Shishido, T. Yamada, A. Thamizhavel, E.
Yamamoto, T. D. Matsuda, Y. Haga, T. Takeuchi, R. Settai, and Y.
Onuki, J. Phys. Soc. Jpn. \textbf{75}, 043703 (2006).

\bibitem{Amano04}
G. Amano, S. Akutagawa, T. Muranaka, Y. Zenitani, and J. Akimitsu,
J. Phys. Soc. Jpn \textbf{73}, 530 (2004).

\bibitem{LiPt-PdB}
K. Togano, P. Badica, Y. Nakamori, S. Orimo, H. Takeya, and K.
Hirata, Phys. Rev. Lett. \textbf{93}, 247004 (2004); P. Badica, T.
Kondo, and K. Togano, J. Phys. Soc. Jpn. \textbf{74}, 1014 (2005).

\bibitem{KOsO}
G. Schuck, S. M. Kazakov, K. Rogacki, N. D. Zhigadlo, and J.
Karpinski, Phys. Rev. B \textbf{73}, 144506 (2006).

\bibitem{MinSam94}
V. P. Mineev and K. V. Samokhin, Zh. Eksp. Teor. Fiz.
\textbf{105}, 747 (1994) [Sov. Phys. JETP \textbf{78}, 401
(1994)].

\bibitem{Agter03}
D. F. Agterberg, Physica C \textbf{387}, 13 (2003).

\bibitem{DF03}
O. V. Dimitrova and M. V. Feigel'man, Pis'ma Zh. Eksp. Teor. Fiz. \textbf{78}, 1132 (2003) [JETP Letters \textbf{78}, 637 (2003)].

\bibitem{Sam04}
K. V. Samokhin, Phys. Rev. B \textbf{70}, 104521 (2004).

\bibitem{MS08}
V. P. Mineev and K. V. Samokhin, Phys. Rev. B \textbf{78}, 144503 (2008).

\bibitem{Lev85}
L. S. Levitov, Yu. V. Nazarov, and G. M. Eliashberg, Pis'ma Zh. Eksp. Teor. Fiz. \textbf{41}, 365 (1985) [JETP Letters
\textbf{41}, 445 (1985)].

\bibitem{Edel89}
V. M. Edelstein, Zh. Eksp. Teor. Fiz. \textbf{95}, 2151 (1989)
[Sov.  Phys.  JETP \textbf{68}, 1244 (1989)].

\bibitem{Yip02}
S. K. Yip, Phys. Rev. B \textbf{65}, 144508 (2002).

\bibitem{Fuji05}
S. Fujimoto, Phys. Rev. B \textbf{72}, 024515 (2005).

\bibitem{GR01}
L. P. Gor'kov and E. I. Rashba, Phys. Rev. Lett. \textbf{87},
037004 (2001).

\bibitem{FAKS04}
P. A. Frigeri, D. F. Agterberg, A. Koga, and M. Sigrist, Phys.
Rev. Lett. \textbf{92}, 097001 (2004) [Erratum \textbf{93},
099903(E) (2004)].

\bibitem{Sam05}
K. V. Samokhin, Phys. Rev. Lett. \textbf{94}, 027004 (2005).

\bibitem{Sam07}
K. V. Samokhin, Phys. Rev. B \textbf{76}, 094516 (2007).

\bibitem{TKKN82}
D. Thouless, M. Kohmoto, M. Nightingale, and M. den Nijs, Phys. Rev. Lett. \textbf{49}, 405 (1982).

\bibitem{Kohm85}
M. Kohmoto, Ann. Phys. (N. Y.) \textbf{160}, 343 (1985).

\bibitem{ON02}
M. Onoda and N. Nagaosa, J. Phys. Soc. Jpn. \textbf{71}, 19 (2002).

\bibitem{JNM02}
T. Jungwirth, Q. Niu, and A. H. MacDonald, Phys. Rev. Lett. \textbf{88}, 207208 (2002).

\bibitem{Hald04}
F. D. M. Haldane, Phys. Rev. Lett. \textbf{93}, 206602 (2004).

\bibitem{Spin-Hall-1}
S. Murakami, N. Nagaosa, and S.-C. Zhang, Science \textbf{301}, 1348 (2003).

\bibitem{Spin-Hall-2}
J. Sinova, D. Culcer, Q. Niu, N. A. Sinitsyn, T. Jungwirth, and A. H. MacDonald, Phys. Rev. Lett. \textbf{92}, 126603 (2004).

\bibitem{KM05}
C. L. Kane and E. J. Mele, Phys. Rev. Lett. \textbf{95}, 146802 (2005); \textit{ibid} \textbf{95}, 226801 (2005).

\bibitem{BZ06}
B. A. Bernevig and S.-C. Zhang, Phys. Rev. Lett. \textbf{96}, 106802 (2006).

\bibitem{FKM07}
L. Fu, C. L. Kane, and E. J. Mele, Phys. Rev. Lett. \textbf{98}, 106803 (2007).

\bibitem{polar-dielec}
R. D. King-Smith and D. Vanderbilt, Phys. Rev. B \textbf{47}, 1651 (1993); R. Resta, Rev. Mod. Phys. \textbf{66}, 899 (1994).

\bibitem{Berry84}
M. V. Berry, Proc. R. Soc. London A \textbf{392}, 45 (1984).

\bibitem{Berry-review}
A. Bohm, A. Mostafazadeh, H. Koizumi, Q. Niu, and J. Zwanziger, \textit{The Geometric Phase in Quantum Systems} (Springer-Verlag, Berlin, 2003).

\bibitem{Zak89}
J. Zak, Phys. Rev. Lett. \textbf{62}, 2747 (1989).

\bibitem{KL54}
R. Karplus and J. M. Luttinger, Phys. Rev. \textbf{95}, 1154 (1954).

\bibitem{CN95}
M. C. Chang and Q. Niu, Phys. Rev. Lett. \textbf{75}, 1348 (1995).

\bibitem{SN99}
G. Sundaram and Q. Niu, Phys. Rev. B \textbf{59}, 14915 (1999).

\bibitem{LL9}
L. D. Landau and E. M. Lifshitz, {\em Statistical Physics, Part 2}
(Butterworth-Heinemann, Oxford, 1980).

\bibitem{Roth66}
L. M. Roth, Phys. Rev. \textbf{145}, 434 (1966).

\bibitem{MS98}
G. P. Mikitik and Yu. V. Sharlai, Zh. Eksp. Teor. Fiz. \textbf{114}, 1375 (1998) [JETP \textbf{87}, 747 (1998)]; Phys. Rev. B \textbf{65}, 184426 (2002).

\bibitem{MS99}
G. P. Mikitik and Yu. V. Sharlai, Phys. Rev. Lett. \textbf{82}, 2147 (1999).

\bibitem{LF91}
R. G. Littlejohn and W. G. Flynn, Phys. Rev. A \textbf{44}, 5239 (1991).

\bibitem{UR85}
K. Ueda and T. M. Rice, Phys. Rev. B \textbf{31}, 7114 (1985).

\bibitem{Rashba60}
E. I. Rashba, Fiz.  Tverd.  Tela (Leningrad) \textbf{2}, 1224
(1960) [Sov. Phys. Solid State \textbf{2}, 1109 (1960)].

\bibitem{Wink-book}
R. Winkler, \textit{Spin-Orbit Coupling Effects in Two-Dimensional Electron and Hole Systems}, Ch. 9 (Springer-Verlag, Berlin, 2003).

\bibitem{SZB04}
K. V. Samokhin, E. S. Zijlstra, and S. K. Bose, Phys. Rev. B \textbf{69}, 094514 (2004) [Erratum: \textbf{70}, 069902(E) (2004)].

\bibitem{LP05}
K.-W. Lee and W. E. Pickett, Phys. Rev. B \textbf{72}, 174505 (2005).

\bibitem{MnSi-exp}
C. Pfleiderer, G. J. McMullan, S. R. Julian, and G. G. Lonzarich, Phys. Rev. B \textbf{55}, 8330 (1997).

\bibitem{BSW-Zak}
L. P. Bouckaert, R. Smoluchowski, and E. Wigner, Phys. Rev. B \textbf{50}, 58 (1936); J. Zak, J. Phys. A: Math Gen. \textbf{35}, 6509 (2002).

\bibitem{DFN85}
B. A. Dubrovin, A. T. Fomenko, and S. P. Novikov, \textit{Modern Geometry -- Methods and Applications. Part II} (Springer-Verlag, New York, 1985).

\bibitem{AshMer}
N. W. Ashcroft and N. D. Mermin, \textit{Solid State Physics} (Brooks/Cole, 1976).

\bibitem{ZFC-review}
\textit{Quantum Mechanics in Phase Space}, ed. C. K. Zachos, D. B. Fairlie, and T. L. Curtright (World Scientific, Singapore, 2005).

\bibitem{MF-book}
V. P. Maslov and M. V. Fedoriuk, \textit{Semiclassical Approximation in Quantum Mechanics} (D. Reidel, Dordrecht, 1981).

\bibitem{Bliokh05}
K. Yu. Bliokh and Yu. P. Bliokh, Ann. Phys. (N. Y.) \textbf{319}, 13 (2005).

\bibitem{PST03}
G. Panati, H. Spohn, and S. Teufel, Comm. Math. Phys. \textbf{242}, 547 (2003).

\bibitem{Vol87}
G. E. Volovik, Pis'ma Zh. Eksp. Teor. Fiz. \textbf{46}, 81 (1987) [JETP Lett. \textbf{46}, 98 (1987)].

\bibitem{BK99}
J. Bolte and S. Keppeler, Ann. Phys. (N.Y.) \textbf{274}, 125 (1999).

\bibitem{CYN05}
D. Culcer, Y. Yao, and Q. Niu, Phys. Rev. B \textbf{72}, 085110 (2005).

\bibitem{Blount62}
E. I. Blount, \textit{Solid State Physics} Vol. 13, ed. H. Ehrenreich, F. Seitz, and D. Turnbull, p. 305 (Academic Press, New York, 1962).

\bibitem{LK04}
I. A. Luk'yanchuk and Y. Kopelevich, Phys. Rev. Lett. \textbf{93}, 166402 (2004).

\bibitem{MS05}
V. P. Mineev and K. V. Samokhin, Phys. Rev. B \textbf{72}, 212504 (2005).

\bibitem{KW02}
S. Keppeler and R. Winkler, Phys. Rev. Lett. \textbf{88}, 046401 (2002).

\bibitem{Sini08}
N. A. Sinitsyn, J. Phys.: Cond. Matter \textbf{20}, 023201 (2008).

\bibitem{NSOMO09}
N. Nagaosa, J. Sinova, S. Onoda, A. H. MacDonald, and N. P. Ong, 
preprint arXiv:0904.4154 (unpublished).

\bibitem{Naga06}
N. Nagaosa, J. Phys. Soc. Jpn. \textbf{75}, 042001 (2006).

\bibitem{QWZ06}
X.-L. Qi, Y.-S. Wu, and S.-C. Zhang, Phys. Rev. B \textbf{74}, 085308 (2006).

\bibitem{Smit55}
J. Smit, Physica \textbf{21}, 877 (1955).

\bibitem{MN03}
S. Murakami and N. Nagaosa, Phys. Rev. Lett. \textbf{90}, 057002 (2003).

\bibitem{SB06}
R. Shindou and L. Balents, Phys. Rev. Lett. \textbf{97}, 216601 (2006).

\bibitem{Messiah}
A. Messiah, \textit{Quantum Mechanics} (Dover Publications, New York, 1999).

\bibitem{ED-Book}
J. P. Elliott and P. G. Dawber, \emph{Symmetry in Physics}, Vol. 2, Ch. 16 (McMillan Press,
London, 1979).

\bibitem{ASS83}
J. E. Avron, R. Seiler, and B. Simon, Phys. Rev. Lett. \textbf{51}, 51 (1983).

\end{thebibliography}
\end{document}